\documentclass[pra,twocolumn,10pt,aps,longbibliography]{revtex4-1}
\usepackage{graphicx}
\usepackage{amsmath} 
\usepackage{amssymb}
\usepackage{amsfonts}
\def\er{{\bf r}}
\def\ek{{\bf k}}

\begin{document}
\title{Dynamical instability of a non-equilibrium exciton-polariton condensate}
\author{Nataliya Bobrovska$^1$, Micha{\l} Matuszewski$^{1}$\footnote{corresponding author, email: mmatu@ifpan.edu.pl}, Konstantinos S. Daskalakis$^2$, Stefan A. Maier$^3$, and St{\'e}phane K{\'e}na-Cohen$^4$}
\affiliation{
$^1$Institute of Physics, Polish Academy of Sciences, Al. Lotnik\'ow 32/46, 02-668 Warsaw, Poland\\
$^2$COMP Centre of Excellence, Department of Applied Physics, Aalto University, PO Box 15100, Fi-00076 Aalto, Finland\\
$^3$Department of Physics, Imperial College London, London SW7 2AZ, United Kingdom\\
$^4$Department of Engineering Physics, Polytechnique Montr{\'e}al, Montr{\'e}al,Qu{\'e}bec H3C 3A7, Canada
}
\begin{abstract}
By imaging single-shot realizations of an organic polariton quantum fluid, we observe the long-sought dynamical instability of non-equilibrium condensates. Without fitting any parameters to our model, we find an excellent agreement between the experimental data and a numerical simulation of the open-dissipative Gross-Pitaevskii equation, which allows us to draw several important conclusions about the physics of the system. We find that the reservoir dynamics are in the strongly nonadiabatic regime, which renders the complex Ginzburg-Landau description invalid. The observed transition from stable to unstable fluid can only be explained by taking into account the specific form of reservoir-mediated instability as well as particle currents induced by the finite extent of the pump spot.
\end{abstract}
\maketitle

\section{Introduction}
Semiconductor microcavities are one of the most versatile systems for realizing and studying quantum fluids of light~\cite{Carusotto_QuantumFluids}. 
As a result of the strong coupling between light and matter modes at resonance, new excitations called exciton-polaritons are formed. These hybrid quasiparticles are coherent superpositions of the semiconductor exciton and microcavity photon, which can inherit properties derived from both fundamental constituents~\cite{Hopfield_Polaritons,Weisbuch_Polaritons,Kavokin_Microcavities}. Although their low effective mass has been touted as an advantage for realizing equilibrium polariton condensates, the driven-dissipative nature of polariton systems--a result of the short particle lifetime--plays an important role in the condensation process. Indeed, the vast majority of experiments are performed in conditions where the polariton gas is out of equilibrium. Still, this allows for several phenomena related to Bose-Einstein condensation to be observed at temperatures much higher than those required for cold atom systems~\cite{Kasprzak_BEC,Yamamoto_PowerLawDecay,Yamamoto_NPReview}, while also stimulating new questions \cite{Altman_DrivenSuperfluid2D}.~ In addition to fundamental research, the physics of non-equilibrium polariton condensates is attractive for its potential applications. Polariton condensates are intrinsically low-threshold sources of coherent light~\cite{Grandjean_RoomTempLasing,Battacharya_RoomTemperatureElectrically} and polariton devices have been used as interferometers~\cite{Bloch_interferometer} and polariton circuit elements~\cite{Bramati_SpinSwitches,Savvidis_TransistorSwitch,Sanvitto_Transistor}.
Moreover, nonlinearities due to strong exciton-mediated interactions give rise to fascinating physical properties such as superfluidity~\cite{Amo_Superfluidity,Deveaud_Vortices}, black hole physics~\cite{Nguyen_AcousticBlackHole} and solitons~\cite{Amo_HydrodynamicSolitons,Santos_BrightSoliton}. Recently, such nonlinearities have also been demonstrated in organic semiconductors~\cite{Kena_Organic,Mahrt_RTCondensatePolymer,KenaCohen_NonlinearOrganic}, which is attractive for the realization of room-temperature devices.

Despite these remarkable developments, the physics of non-resonantly pumped polariton condensates is still not completely understood. Condensation requires external optical~\cite{Kasprzak_BEC} or electronic~\cite{Hofling_ElectricallyPumped,Battacharya_RoomTemperatureElectrically} pumping, which creates a reservoir of high-energy electronic excitations. The energetic relaxation of these excitations due to interaction with the environment enhanced by bosonic stimulation can then lead to a macroscopic occupation of the low-lying polariton ground state. 
This complicated process has been modeled theoretically within various approximations~\cite{Haug_QuantumKineticGPDerivation,Wouters_ClassicalFields,Malpuech_Hybrid,Laussy_SpontaneousCoherence,Bloch_Molecules,Tassone_Bottleneck}. A particularly useful description is based on the phenomenological open-dissipative Gross-Pitaevskii equation (ODGPE)~\cite{Wouters_ExcitationSpectrum}. Its application is widespread due to the simplicity of this description, the limited number of free parameters required and its success in reproducing experimental results. We note that a similar model was also used to describe non-equilibrium condensates in atom laser systems~\cite{Kneer_GenericAtomLaser}.

Since the introduction of the ODGPE model~\cite{Wouters_ExcitationSpectrum} it was realized that for certain parameters it predicts a peculiar instability of the condensate due to the interaction of polaritons with the reservoir of uncondensed excitons. This dynamical instability is due to the repulsive interaction of condensed polaritons with reservoir excitons, and is excpected to result in phase separation of the two components. Since to date there was no experimental evidence of this instability, its physical relevance was unclear. Some authors have suggested that the instability is an artifact that would disappear when energy relaxation in the condensate was properly accounted for~\cite{Malpuech_Hybrid,Wouters_EnergyRelaxation, Carusotto_NonequilibriumQuasicondensates}. Many theoretical studies circumvented this problem by imposing the adiabatic assumption of a fast reservoir response, either indirectly by choosing a reservoir lifetime much shorter than typical lifetime of an exciton, or directly by using a simplified complex Ginzburg-Landau (CGLE) description with no separate reservoir degree of freedom~\cite{Keeling_VortexDynamics, Sieberer_DynamicalCritical}. In this regime, the model becomes instability free~\cite{Bobrovska_Adiabatic}.

Here, we demonstrate that the reservoir-induced instability is a real phenomenon that can occur in non-equilibrium polariton condensates. We confirm this by measuring single-shot realizations of the condensate emission from an oligofluorene-filled organic microcavity. The results are compared to the predictions of the ODGPE model without any parameter fitting. Previous measurements of first-order spatial correlations in this system hinted at the possible breakdown of the stable condensate model~\cite{Kena_SpatialCoherence}. Excellent agreement between experiment and numerical modeling allows us to determine that the lifetime of the reservoir indeed places the system in the unstable and strongly nonadiabatic regime, where the simplified CGLE-like description does not provide a reliable description of system dynamics. We also demonstrate that there is a transition from a stable to an unstable condensate with increasing pump power. This behavior is in opposition with the previously reported stability criterion obtained for continuous wave pumping~\cite{Ostrovskaya_DarkSoliton}. We explain this seeming contradiction as resulting from a competition between reservoir-induced instability, finite condensate lifetime under pulsed excitation, and stabilizing effect of particle currents. Although the experiment we describe was performed on an organic microcavity, the ODGPE model we use to interpret the results holds equally well for inorganics (neglecting the spin degree of freedom). We conclude by discussing the relevance of our findings to the stability and coherence of inorganic polariton condensates.

\section{Model}
We model the exciton-polariton condensate using the two-dimensional stochastic ODGPE 
for the wavefunction  $\psi(\er,t)$ coupled to
the rate equation for the polariton reservoir density, $n_R(\er,t)$ 
\cite{Wouters_ExcitationSpectrum,Wouters_ClassicalFields}
\begin{align}
\label{GPE-psi}
i\textrm{d}\psi=&\left[ -\frac{\hbar}{2m^*}\nabla^2+\frac{g_C}{\hbar}|\psi|^2+\frac{g_R}{\hbar}n_R+\right.\\\nonumber&\left.
+\frac{i}{2}\left(Rn_R-\gamma_C\right)
\right]\psi\textrm{d}t+\textrm{d}W,\\
\label{GPE-nr}
\frac{\partial n_R}{\partial t}=&P-\left(\gamma_R+R|\psi|^2\right)n_R-k_b n_{R} ^2,
\end{align}
where $P(\er,t)$ is the exciton creation rate due to the pumping pulse, $m^*$ 
is the effective mass of lower polaritons, 
$\gamma_C$ and $\gamma_R$ are the polariton and exciton dissipation rates, 
$R$ is the rate of stimulated scattering to the condensate, $g_C$ an $g_R$ are the polariton-polariton and polariton-reservoir interaction coefficients, respectively, and $k_b$ is the bimolecular annihilation rate. The latter is specific property of organic semiconductors, but it does not qualitatively affect the calculation results. With the exception of the pumping and dissipation terms, Eq.~\ref{GPE-psi} is analogous to the Gross-Pitaevskii equation used to describe atomic condensates, where $g_C$ plays the role of a contact interaction. The quantum noise d$W$ can be obtained within the truncated Wigner approximation~\cite{Wouters_ClassicalFields} as Gaussian noise with correlations
$\langle\textrm{d}W({\bf r})\textrm{d}W^*({\bf r}^{\prime})\rangle=\frac{\textrm{d}t}{2(\Delta  x)^2}(Rn_R+\gamma_C)\delta_{{\bf r},{\bf r}^{\prime}}$ and
$\langle\textrm{d}W({\bf r})\textrm{d}W({\bf r}^{\prime})\rangle=0$
where $\Delta x$ is the lattice constant of the discretized mesh.

\begin{figure}
\centering
\includegraphics[width=\linewidth]{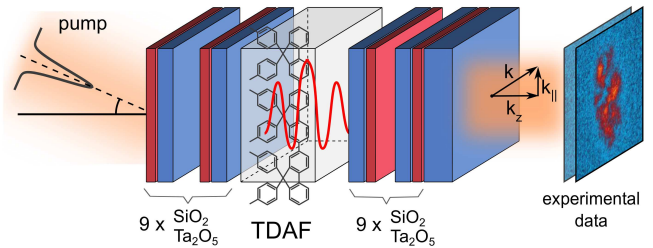}
\caption{Schematic of the sample, which is composed of an amorphous oligofluorene film sandwiched between two dielectric Bragg mirrors. A high energy impulsive pump was incident on the sample at a $\theta=50^\circ$ angle. The near-field images, representatively shown on the right, were obtained by collecting the photoluminescence on a CCD camera.}
\label{fig:scheme}
\end{figure}

The cavity under consideration is shown schematically in Fig.~\ref{fig:scheme} and composed of a single thin film of 2,7-bis[9,9-di(4-methylphenyl)-fluoren-2-yl]-9,9-di(4-methylphenyl)fluorene (TDAF) sandwiched between two dielectric Bragg mirrors composed 9 pairs of alternating Ta$_2$O$_5$/SiO$_2$. For each laser shot, a time-integrated image of the condensate emission was recorded using a CCD camera (see Methods for details). All of the parameters were obtained independently in previous measurements using a different technique, and are summarized in the Methods Section. In Eq.~\eqref{GPE-nr}, we assume that the pump pulse length ($\tau_{pulse}=250$ fs) is short enough that $P(\er,t)$ as a $\delta(t)$ function in time. We neglect the effect of static disorder of the sample. This assumption will be justified by comparison with experimental data.

\section{Instability of the condensate}
As previously reported, the condensate solution is prone to dynamical instabilities for a certain range of parameters~\cite{Wouters_ExcitationSpectrum}. The physical origin of this instability is the repulsive interaction $g_R$ between the condensate and reservoir excitons, which can lead to phase separation of these two components. For the parameters of our system, the instability is predicted to occur for all continuous pump powers below $P=(g_R \gamma_C)/(g_C \gamma_R)P_{\rm th}\approx 3000 P_{\rm th}$~\cite{Ostrovskaya_DarkSoliton}. This greatly exceeds the range of pump powers accessible here and in any other organic microcavity for typical values $\gamma_C$ and $\gamma_R$. Many inorganic microcavities also fall within the instability regime.  

\begin{figure*}
\centering
\includegraphics[width=1.\linewidth]{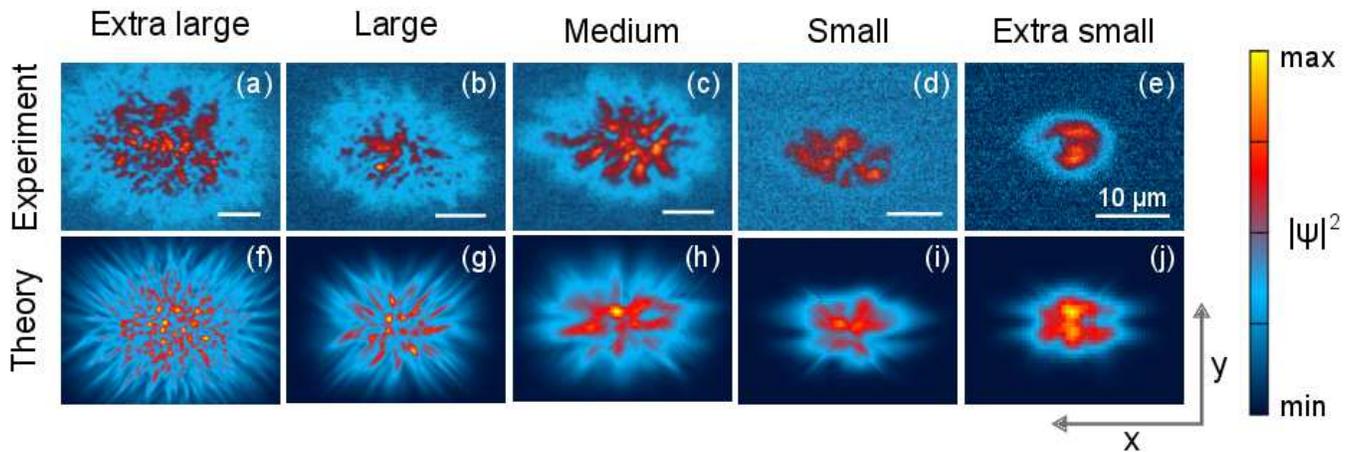}
\caption{Comparison between the time-integrated experimental (top) and numerical (bottom) polariton field density. The size of the pump spot decreases from (a), (f) to (e), (j). The instability leads to the creation of polariton domains. The color scale is normalized to maximum value in each frame separately. Parameters used in numerical simulations are $m^*=2.1\times 10^{-5} m_e$, $R= 1.1 \times 10^{-2}\,$cm$^2$s$^{-1}$, $\gamma_C= (167\,{\rm fs})^{-1}$, $\gamma_{\rm R}$=$(300\,{\rm ps})^{-1}$, $g_C=10^{-6}$ meV $\mu$m$^2$, $g_R=1.7\times 10^{-6}$ meV $\mu$m$^2$, $k_b=3.3 \times 10^{-5} {\rm cm^2s^{-1}}$ and correspond to the values measured in~\cite{KenaCohen_NonlinearOrganic}.}
\label{fig:profiles}
\end{figure*}

Single-shot real space images of the condensate photoluminescence are shown in Fig.~\ref{fig:profiles}(a)-(e) for varying dimensions of the Gaussian pump. These were taken at a pump power $P=2P_{th}$ for Fig.~\ref{fig:profiles}(a)-(d) and $P=2.5 P_{th}$ for Fig.~\ref{fig:profiles}(e), where $P_{th}$ is the condensation threshold. The corresponding ODGPE calculations for the same powers, where only the shape of the pump is varied are shown below in Fig.~\ref{fig:profiles}(f)-(j). The agreement is remarkable given that no parameter fitting was applied. The experiment and calculation highlight that the instability is more pronounced for large spatial pump sizes (or flat-top, which is not shown). In contrast, the smallest condensate size is only slightly affected by the instability.

\begin{figure}
\centering
\includegraphics[width=\linewidth]{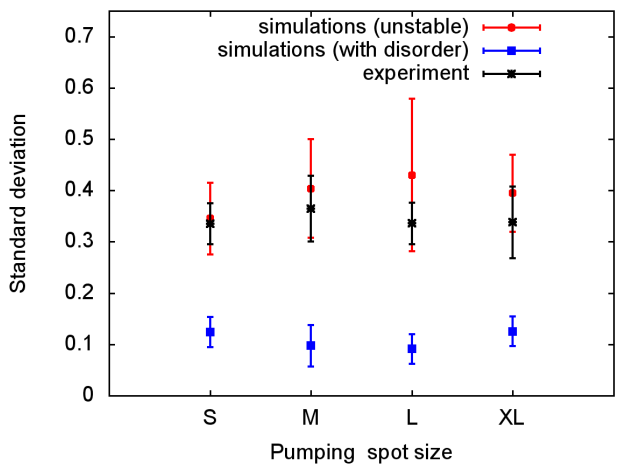}
\caption{Effect of sample disorder. Normalized standard deviation of shot-to-shot local luminescence $\sigma[I(\er_0)]/\langle I(\er_0)\rangle$ as measured int the experiment, and predicted by the model in (a) the unstable regime and (b) in the stable regime with the addition of static disorder (40 shots in each case). The disorder amplitude was chosen to be large enough so that its presence can induce the creation of polariton domains, similar as in the experiment, and its correlation length was tuned to the typical domain size. The error bars are obtained by averaging over 5 different points on the sample $\er_0$.}
\label{fig:stddev}
\end{figure}

Note that both the experiment and calculation are time-averaged over the $duration$ of the condensate emission for each pulse.  The exact size and orientation of the patterns varies randomly from shot to shot both in experiment and simulations, but these remain qualitatively the same (see Appendix~\ref{app:single-shot}). This suggests that disorder does not play an important role in determining the final condensate profile. 
To further verify this assumption quantitatively, we measured the shot-to-shot variations of the luminescence intensity from several points on the sample. The experimental results are compared on Fig.~\ref{fig:stddev} with numerical results in two cases: the unstable regime of parameters, and the stable regime with the addition of a static disorder potential $V(\er)$. The variation of intensity is in the latter case much lower than observed experimentally, due to pinning of domains to the potential minima. This allows to discard the hyphotesis that domains are created in a stable condensate solely due to the effect of disorder.

The excellent agreement between the experiment and theory allows us to draw some important conclusions about the physics of the system. The parameters of the model indicate that the dynamics are not only in the unstable, but also strongly {\it nonadiabatic} regime, i.e.~the reservoir $n_R(\er,t)$ does not quickly follow the changes in the condensate density $|\psi(\er,t)|^2$. The adiabatic regime is attained only when three independent analytical conditions are fulfilled simultaneously (see Appendix~\ref{app:adiabaticity}). Here, all three conditions are violated. In particular $g_R n_R \approx 40 (\gamma_R + R |\psi|^2)$, which means that the reaction time of the reservoir is 40 times slower than the response time of the condensate due to interactions with the reservoir. 

\begin{figure}
\centering
\includegraphics[width=0.8\linewidth]{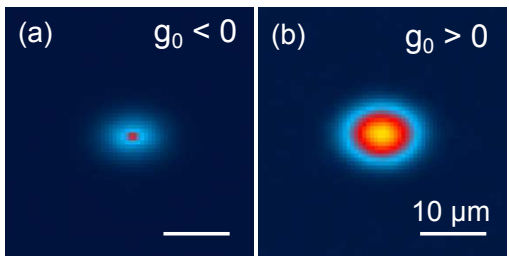}
\caption{Results of numerical simulations using the CGLE-based stochastic Gross-Pitaevskii equation. Parameters correspond to the extra small spot (e) case from Fig.~\ref{fig:profiles} (see text).}
\label{fig:sgpe}
\end{figure}

The breakdown of the adiabatic approximation suggests that CGLE-like models based on a single equation cannot reliably describe the dynamics of the system. This is due to the fact that they do not not incorporate the reservoir as a separate degree of freedom. We demonstrate this by numerically modeling the single stochastic Gross-Pitaevskii equation (SGPE)~\cite{Carusotto_NonequilibriumQuasicondensates,Wouters_SpatialCoherence} for the same parameters. As shown in Fig.~\ref{fig:sgpe}(a), this model gives a poor agreement to the experimental data. The coefficients of the SGPE equation were determined using the correspondence formulas derived in~\cite{Bobrovska_Adiabatic}. The effective interaction between polaritons turns out to be attractive, which is due to the reservoir-mediated attraction in the unstable regime. Nevertheless, the instability does not give rise to multiple domains, but rather to condensate collapse with no spatial symmetry breaking.
We also verified that SGPE simulations with an explicit repulsive interaction coefficient are not able to reproduce the experimental patterns, see Fig.~\ref{fig:sgpe}(b). The details of the model are given in the Supplementary Information.

\section{Transition from stable to unstable condensate}
For a large size of the pump spot, we generally observe the instability independently of the pump pulse power, see Fig.~\ref{fig:profiles}. However, in the case of a small size (eg.~Fig.~\ref{fig:profiles}(e,j), a transition from a stable to an unstable condensate is seen with increasing pump power (see Supplementary Fig.~1). When the power of the pump pulse is less than about $1.8P_{th}$, a single condensate is formed, while for pump powers above this value, the instability results in appearance of two or more domains. Note that this stable region is precisely where first-order spatial coherence measurements were previously reported. Stability is a necessity due the extraction procedure, which requires fitting several interferograms as a function of phase delay. Any shot-to-shot fluctuations consequently lead to an artificial reduction of the fringe visibility. Finally, via numerical simulations, we observe that the transition from stable to unstable is not abrupt and that there is some shot-to-shot variation along the boundary.

The observed power dependence is in contradiction with the previously predicted transition from an unstable to a stable condensate with increasing continuous wave pumping~\cite{Ostrovskaya_DarkSoliton,Liew_InstabilityInduced}. To understand this effect we developed a theoretical model of condensation dynamics under impulsive excitation. The main elements that determine the stability of the condensate are the unstable dynamical Bogoliubov spectrum the condensate, the density current of polaritons flowing away from the center of the pumping spot, and the finite lifetime of the condensate. The competition of these processes can explain the existence of stable or unstable condensate for various pumping conditions, and allows for the calculation of the transition region.

 \begin{figure*}
 \centering
 \includegraphics[width=\linewidth]{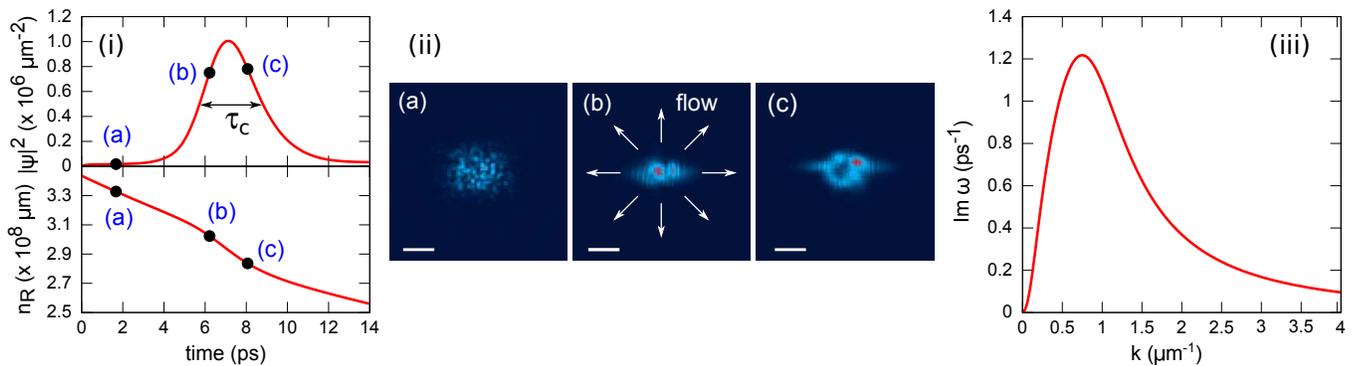}
 \caption{The development of the condensate instability for the small Gaussian spot at power $P=1.6P_{th}$. The left panel (i) shows the simulated evolution of the condensate and reservoir particle number, with the condensate lifetime $\tau_C$ defined as the FWHM duration of the condensate emission.  Panel (ii) with frames (a)-(c) shows snapshots of density profiles at the three chosen instants of time. The outgoing flow of polaritons is responsible for the creation of a single condensate. At high pumping, the Bogoliubov instability breaks up the condensate, as shown in frames (b)-(c). (iii) Imaginary part of the frequency of the unstable Bogoliubov branch calculated with (\ref{matrix1}) at peak polariton density.}
 \label{fig:evolution}
 \end{figure*}

 The evolution of the system can be divided into several stages shown in Fig.~\ref{fig:evolution}(a)-(c).
At the arrival of the ultrashort pulse, the reservoir density is fixed according to the Gaussian profile of the pump, while the wavefunction $\psi$ contains only random fluctuations inherited from the Wigner noise. Due to the finite size of the pump spot, a polariton current from the center of the pump spot is created by the reservoir-induced potential. This favors the formation of a single condensate, as the most central fluctuation ``spills over'' and repels the other domains outside. 
However, for large pumping powers a single condensate is not formed, and the system evolves to a fragmented state, while some flow of polaritons from the center is still visible. We attribute this effect to the existence of unstable Bogoliubov modes which break up the condensate. These unstable modes are always present, but for low pumping powers the instability is too weak to develop during the lifetime of the condensate. The condensate lifetime is understood here as the FWHM duration of the emission from the condensate, see Fig.~\ref{fig:evolution}.

To quantify this, we have calculated the Bogoliubov instability timescale in the local-density approximation, i.e.~neglecting the spatial inhomogeneity of the pump. We find that in the case of a pulsed pump, the dynamical Bogoliubov spectrum can be qualitatively different from the one in the case of
continuous pumping, which was considered in previous reports~\cite{Wouters_ExcitationSpectrum,Byrnes_Yamamoto_NegativeBogoliubov,Ostrovskaya_DarkSoliton, Bobrovska_Stability},
due to the absence of the pumping term after the arrival of the pulse (see Methods for details).
For our parameters, the imaginary part of the unstable Bogoliubov branch frequency is shown in Fig.~\ref{fig:evolution}(iii) and is inversely related to the instability timescale.

\begin{figure}
\centering
\includegraphics[width=\linewidth]{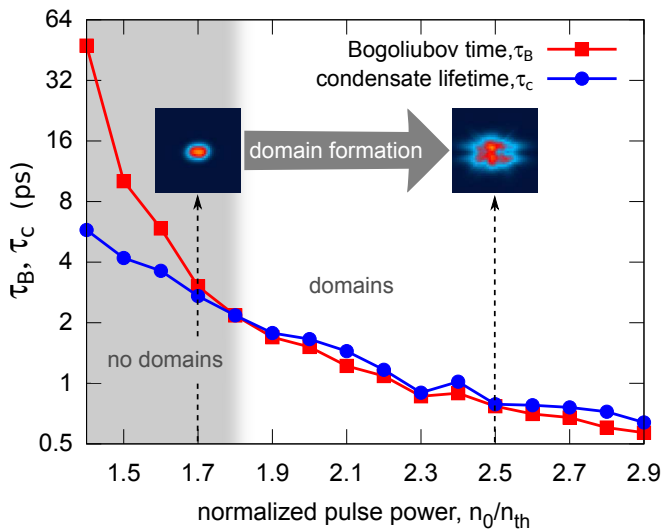}
\caption{Comparison between the characteristic time scales that are responsible for the development of the instability. According to numerical simulations, polariton domains are formed only when the Bogoliubov instability time $\tau_{\rm B}$ becomes comparable to the lifetime of the condensate emission $\tau_{\rm C}$, which is determined as in Fig.~\ref{fig:evolution}. At lower density, a single symmetric condensate is created. Averaged over 5 realizations.}
\label{fig:timescales}
\end{figure}

A comparison between the Bogoliubov instability timescale and the condensate lifetime is shown in Fig.~\ref{fig:timescales}. We find that the Bogoliubov timescale $\tau_{B}$ is longer than the condensate lifetime $\tau_{C}$ for $n_0/n_{th} \lesssim 1.8$, where $n_0$ is the maximum of $n_R(\er,t=0)$ and $n_{th}$ is the threshold value of $n_0$ for condensate formation, see Fig.~\ref{fig:timescales}. Above this value these timescales become comparable. This is in very good agreement with the observed threshold for domain formation, see Supplementary Figure~1. The similarity between the timescales $\tau_{B}$ and $\tau_{C}$ above $n_0/n_{th} \approx 1.8$ is explained by the similar magnitude of all nonlinear coefficients ($\hbar R$, $g_C$, and $g_R$) in Eq.~(\ref{GPE-psi}). At high pump powers, the maximum density of the condensate $|\psi|^2$ becomes comparable to $n_R$, and all nonlinear energy scales have similar order of magnitude; in particular, the spontaneous scattering rate $R |\psi|^2$, which depletes the reservoir and influences the lifetime.




\section{Relevance to inorganic condensates}
The vast majority of exciton-polariton condensates are realized in inorganic semiconductors, where properties are slightly different from the organic case considered here.
Nevertheless, the parameters of inorganic samples also place them in the unstable regime.
In particular, independent measurements~\cite{Bloch_DynamicsElectronGas} as well as modeling of dynamics~\cite{Bloch_PropagationAmplification,Gippius_Bistability,Deveaud_Josephson} indicate a reservoir lifetime in the hundreds of picoseconds, which suggests that it should not be treated adiabatically. In several experiments, however, the "bottleneck" region plays the role of the reservoir and the relaxation kinetics may need to be considered. In contrast, organic microcavities have short enough polariton lifetimes that single-step relaxation processes from the reservoir can be considered to be dominant. Meanwhile, instabilities in inorganic condensates may be present in some systems, but obscured because of the absence of single-shot measurements. The difficulty in performing such measurements is mostly due to the lower polariton densities typical of inorganic microcavities. In our samples, averaging over tens of pulses already washes out clear signatures of the domain formation. An alternative to single-shot experiments is the measurement of spatial correlation functions in which the signatures of domains can persist. Indeed, this is observed in the microcavity considered here as a reduction in the first-order spatial coherence~\cite{Kena_SpatialCoherence}. 

\section{Conclusions}

In conclusion, we have demonstrated for the first time the reservoir-mediated instability of a non-equilibrium exciton-polariton condensate. Excellent agreement between the experiment and theory suggests that models with reservoir treated as a separate degree of freedom should be used to describe these systems. Under pulsed excitation, we find that various timescales determine that stability limit, including the finite condensate duration, the reservoir-induced instability and particle currents due to repulsive exciton-polariton interactions.

\acknowledgments NB and MM acknowledge support from the National Science Center of Poland grants DEC-2011/01/D/ST3/00482 and 2015/17/B/ST3/02273. SKC acknowledges funding from the NSERC Discovery Grant program. SAM acknowledges the EPSRC Active Plasmonics Programme EP/H000917/2, the Royal Society, and the Lee-Lucas Chair in Physics. KSD acknowledges the Leverhulme Trust and EPSRC Active Plasmonics Programme.

\appendix
  
\section{Methods} \label{app:methods}

The sample fabrication was previously described in Ref.~\cite{KenaCohen_NonlinearOrganic}. Samples were fabricated on quartz substrates and excitation and detection were performed in a transmission geometry. The pulsed excitation was provided by an optical parametric amplifier (Orpheus, Light Conversion) pumped by an Yb:KGW amplifier operating at 100 Hz (PHAROS, Light Conversion). The slow repetition rate allowed sufficient time for the CCD camera (Thorlabs BC106-VIS) to capture each frame within the read time. 

Simulations of the condensate dynamics were performed using a fourth-order Runge Kutta algorithm with absorbing boundary conditions.

\section{Randomness of single-shot density patterns and the role of disorder}  \label{app:single-shot}

\begin{figure}
 \centering
 \includegraphics[width=\linewidth]{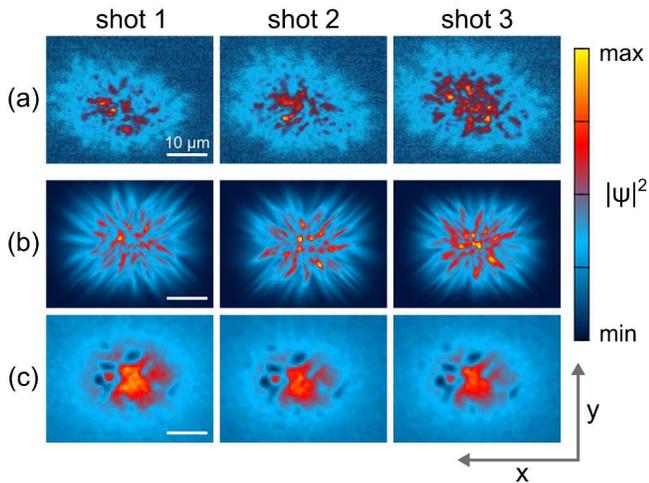}
 
 \caption{Examples of single-shot luminescence patterns from experiments (a), numerical simulations in the unstable regime (b),
   and numerical simulations in the stable regime, where the polariton domains are created by disorder (c). The size of the pump corresponds to the ``Large'' spot, and other parameters are as in Figure~\ref{fig:profiles}.}
 \label{fig:single-shots}
 \end{figure}

In Figure~\ref{fig:single-shots} we show examples of single-shot luminescence patterns, averaged over duration of emission after a single pulse excitation. The experimental results (a) are compared with numerical results in two cases: the unstable regime of parameters, as in Fig.~\ref{fig:profiles}, and stable regime with the addition of a disorder $V(\er)$ (b). The stable regime was obtained here by the reduction of the pump power, as in Fig.~\ref{fig:timescales}. The disorder amplitude was chosen to be large enough so that its presence can induce the creation of polariton domains similar as in the experiment, and its correlation length was tuned to the typical domain size.

Clearly, the disorder-induced domains in the stable regime (c) are placed in the same positions from shot to shot, while both in experiments and simulations (b) their placement is random. This effect leads to the observed smearing out of the domain patterns when averaging over multiple shots. This is in contrast to previous works, where polariton domains were observed in luminescence averaged over a long time or multiple shots (eg.~\cite{Kasprzak_BEC,Deveaud_VortexDynamics,Deveaud_Disorder}), pointing out that their position was fixed by local disorder fluctuations.

\section{Adiabaticity conditions} \label{app:adiabaticity}
  
As demonstrated in~\cite{Bobrovska_Adiabatic}, the correspondence between the ODGPE and CGLE or SGPE models holds in the continuous pumping case when several conditions are fulfilled simultaneously. The first condition is that the fluctuations around the steady state are small, and the other are given by the formulas
\begin{equation} \label{condk}
k^2 \ll 2m^*/(\hbar\tau_R)
\end{equation}
where $\tau_R=(\gamma_R + R |\psi({\bf r},t)|^2)^{-1}$ and $k$ is the maximum momentum that is relevant for the dynamics, and
\begin{align}
\frac{P_{\rm th}}{P}\gg\frac{g_C-R}{g_C},\label{cond1}\\
\frac{P}{P_{\rm th}}\gg\frac{g_R}{R}\frac{\gamma_C}{\gamma_R}.\label{cond2}
\end{align}

\section{Pulsed Bogoliubov spectrum and the role of polariton-polariton interactions}

Instead of examining stability about a steady-state, we consider small fluctuations around the homogeneous state with a polariton and reservoir density evolving in time 
\begin{flalign} 
\label{bogo-uv}
\psi&=\psi_0 e^{-\frac{i\mu t}{\hbar}+\frac{\beta_Ct}{2}}\Bigg( 1 +\\ \nonumber&
+ \sum_{\ek} \left[ u_\ek e^{-i(\omega_\ek t-\ek \er)}+v_\ek ^* e^{i(\omega_\ek ^*t-\ek \er)}\right]\Bigg),\\
n_R&=n_R ^0 e^{\beta_Rt} \left(1
+ \sum_{\ek} \left[ w_\ek e^{-i(\omega_\ek t-\ek \er)}+ \mathrm{c.c.}
  \right]\right), \nonumber
\end{flalign}
where $\omega_\ek$ is the frequency of the mode with the wavenumber $\ek$, and $u_\ek,v_\ek,w_\ek$ are small fluctuations.
In the above $\beta_{C,R}$ are condensate and reservoir density growth or decay rates, that need to be taken into account in the case of pulsed excitation.

Consider a condensate and reservoir that are approximately spatially homogeneous (as in the standard Bogoliubov method) but with the average densities increasing or decreasing in time due to imbalance in loss and gain. Small spatial fluctuations on top of this state may grow or decay independently of the growth or decay of the average condensate or reservoir density. For example, while the average condensate density grows from $t_1$ to $t_2$ by 10\%, the fluctuations of the wavefunction in space may grow from 1\% of the average amplitude at $t_1$ to 2\% of average amplitude at $t_2$, i.e. the condensate wavefunction may become more ``rough''. When $t_1$ and $t_2$ differ by an infinitesimally small amount $dt$, we can calculate the instantaneous growth rate of these fluctuations analytically. In practice, we can calculate the complex frequency of these fluctuations. 

The growth rates calculated in this way are time-dependent, since the state of the condensate and reservoir change in time. After long time of evolution, growth (or decay) of a fluctuation will be given by its instantaneous growth rate integrated over time.
We note that we treat the condensate and reservoir (both the homogeneous base states and fluctuations around them) as separate degrees of freedom, therefore the calculation is not limited to the adiabatic approximation. The resulting fluctuation eigenmodes are appropriate combinations of condensate and reservoir fluctuations.

The excitation spectrum is described by the eigenvalue problem $\mathcal{L}_\ek \mathcal{U}_\ek=\hbar\omega_\ek \mathcal{U}_\ek$, where ${\mathcal{U}_\ek=(u_\ek, v_\ek, w_\ek)^T}$ and
\begin{align}
 \label{matrix1}
&\mathcal{L}_\ek =
\begin{pmatrix}
  g_C n_p+\epsilon_k & g_C n_p & \left(\frac{i\hbar}{2}R^{2D}+g_R\right)n_R^0\\
  -g_C n_p & -g_C n_p-\epsilon_k &  \left(\frac{i\hbar}{2}R^{2D}-g_R\right)n_R^0 \\
 -i\hbar R^{2D} n_p & -i\hbar R^{2D} n_p  & -i\hbar (P/n_R^0+k_b n_R^0)
 \end{pmatrix}
\end{align}
and $\epsilon_k=\hbar^2k^2/2m^*$, $n_p=|\psi_0|^2$, $\mu = n_p g_C+g_R n_R^0$, $\beta_C = R^{2D} n_R^0-\gamma_C$, $\beta_R = P/n_R^0 -n_p R^{2D} -\gamma_R$. In the special case of a stationary state under CW (continuous wave) pumping
the above matrix is equivalent to the one considered before~\cite{Wouters_ExcitationSpectrum}.

In the pulsed case, we have $P=0$ after the arrival of the pulse,
and consequently $\beta_{R}<0$, i.e.~the decay of the reservoir density. It is easy to see that the matrix~(\ref{matrix1}) has then a qualitatively different form, and the calculated spectrum differs substantially from the CW case.

\begin{figure}[tbp]
 \centering
 \includegraphics[width=\linewidth]{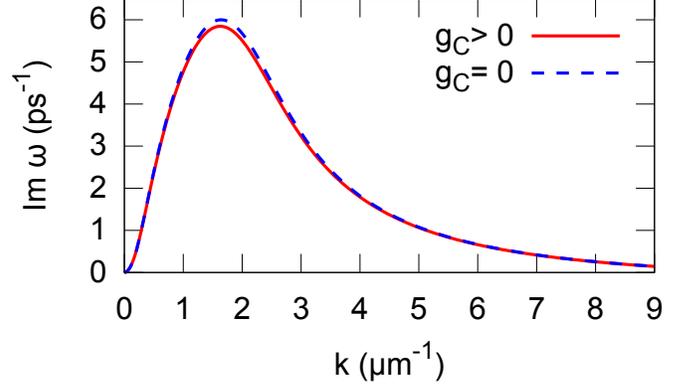}
 \caption{Bogoliubov spectrum with and without polariton-polariton interaction term, $g_C$. Parameters correspond to the small Gaussian pumping spot at power $P=2.0P_{th}$.}
 \label{fig:bogo-wigner}
 \end{figure}

The Bogoliubov instability time $\tau_{B}$ is calculated from the Bogoliubov spectrum taking
the polariton density equal to half of the maximum density (corresponding to the spatial average
with a Gaussian shape of the condensate)
times one-half due to the temporal dependence of the density as in Fig.~\ref{fig:evolution}. For consistency,
the condensate lifetime is taken as the full width at half maximum of the temporal dependence of the density.
The instability time scale is estimated as $t_{\rm B}=2.5/{\rm max}_k ({\rm Im} \omega_k)$, where we assume that initial density fluctuations
on the level of $8\%$ of the maximum density can develop into domains. These fluctuations are not mainly seeded by the
Wigner noise, which are rather small, but by the inhomogeneous shape of the reservoir. The value of $8\%$ fluctuation
is consistent with the spatial variation of the initial reservoir density on the area where the condensate is typically formed.

We emphasize that the flow of polaritons explains the formation of a single domain from initial noise at low pumping powers, as follows from analysis of numerical simulations.  This process is a linear effect, independent of the density, so it is not limited by the lifetime of the high-density condensate, but by the time of the formation of the condensate. For this reason it dominates at low pumping powers.

The role of polariton-polariton interactions can be estimated by artificially turning off the $g_C$ interaction term in the Gross-Pitaevskii equation. We find that this term has little influence on both the Bogoliubov spectrum and the dynamics of the condensate, see Figure \ref{fig:bogo-wigner}. The instability is only slightly stronger without the polariton interaction term. Therefore, we conclude that the instability is solely due to the polariton-reservoir repulsive interaction $g_R$.

\bibliography{references_PRX}

\begin{thebibliography}{10}%
\makeatletter
\providecommand \@ifxundefined [1]{%
 \ifx #1\undefined \expandafter \@firstoftwo
 \else \expandafter \@secondoftwo
\fi
}%
\providecommand \@ifnum [1]{%
 \ifnum #1\expandafter \@firstoftwo
 \else \expandafter \@secondoftwo
\fi
}%
\providecommand \enquote [1]{``#1''}%
\providecommand \bibnamefont  [1]{#1}%
\providecommand \bibfnamefont [1]{#1}%
\providecommand \citenamefont [1]{#1}%
\providecommand\href[0]{\@sanitize\@href}%
\providecommand\@href[1]{\endgroup\@@startlink{#1}\endgroup\@@href}%
\providecommand\@@href[1]{#1\@@endlink}%
\providecommand \@sanitize [0]{\begingroup\catcode`\&12\catcode`\#12\relax}%
\@ifxundefined \pdfoutput {\@firstoftwo}{%
 \@ifnum{\z@=\pdfoutput}{\@firstoftwo}{\@secondoftwo}%
}{%
 \providecommand\@@startlink[1]{\leavevmode\special{html:<a href="#1">}}%
 \providecommand\@@endlink[0]{\special{html:</a>}}%
}{%
 \providecommand\@@startlink[1]{%
  \leavevmode
  \pdfstartlink
   attr{/Border[0 0 1 ]/H/I/C[0 1 1]}%
   user{/Subtype/Link/A<</Type/Action/S/URI/URI(#1)>>}%
  \relax
 }%
 \providecommand\@@endlink[0]{\pdfendlink}%
}%
\providecommand \url  [0]{\begingroup\@sanitize \@url }%
\providecommand \@url [1]{\endgroup\@href {#1}{\urlprefix}}%
\providecommand \urlprefix [0]{URL }%
\providecommand \Eprint[0]{\href }%
\@ifxundefined \urlstyle {%
  \providecommand \doi [1]{doi:\discretionary{}{}{}#1}%
}{%
  \providecommand \doi [0]{doi:\discretionary{}{}{}\begingroup
  \urlstyle{rm}\Url }%
}%
\providecommand \doibase [0]{http://dx.doi.org/}%
\providecommand \Doi[1]{\href{\doibase#1}}%
\providecommand \bibAnnote [3]{%
  \BibitemShut{#1}%
  \begin{quotation}\noindent
    \textsc{Key:}\ #2\\\textsc{Annotation:}\ #3%
  \end{quotation}%
}%
\providecommand \bibAnnoteFile [2]{%
  \IfFileExists{#2}{\bibAnnote {#1} {#2} {\input{#2}}}{}%
}%
\providecommand \typeout [0]{\immediate \write \m@ne }%
\providecommand \selectlanguage [0]{\@gobble}%
\providecommand \bibinfo [0]{\@secondoftwo}%
\providecommand \bibfield [0]{\@secondoftwo}%
\providecommand \translation [1]{[#1]}%
\providecommand \BibitemOpen[0]{}%
\providecommand \bibitemStop [0]{}%
\providecommand \bibitemNoStop [0]{.\EOS\space}%
\providecommand \EOS [0]{\spacefactor3000\relax}%
\providecommand \BibitemShut [1]{\csname bibitem#1\endcsname}%
\bibitem{Carusotto_QuantumFluids}%
  \BibitemOpen
  \bibfield{author}{%
  \bibinfo {author} {\bibfnamefont{Iacopo}\ \bibnamefont{Carusotto}}\ and\
  \bibinfo {author} {\bibfnamefont{Cristiano}\ \bibnamefont{Ciuti}},\ }%
  \bibfield{title}{%
  \enquote{\bibinfo {title} {Quantum fluids of light},}\ }%
  \bibfield{journal}{%
  \Doi{10.1103/RevModPhys.85.299}{\bibinfo {journal} {Rev. Mod. Phys.}}\ }%
  \textbf{\bibinfo {volume} {85}},\ \bibinfo {pages} {299--366} (\bibinfo
  {year} {2013})%
  \bibAnnoteFile{NoStop}{Carusotto_QuantumFluids}%
\bibitem{Hopfield_Polaritons}%
  \BibitemOpen
  \bibfield{author}{%
  \bibinfo {author} {\bibfnamefont{J.~J.}\ \bibnamefont{Hopfield}},\ }%
  \bibfield{title}{%
  \enquote{\bibinfo {title} {Theory of the contribution of excitons to the
  complex dielectric constant of crystals},}\ }%
  \bibfield{journal}{%
  \Doi{10.1103/PhysRev.112.1555}{\bibinfo {journal} {Phys. Rev.}}\ }%
  \textbf{\bibinfo {volume} {112}},\ \bibinfo {pages} {1555--1567} (\bibinfo
  {year} {1958})%
  \bibAnnoteFile{NoStop}{Hopfield_Polaritons}%
\bibitem{Weisbuch_Polaritons}%
  \BibitemOpen
  \bibfield{author}{%
  \bibinfo {author} {\bibfnamefont{C.}~\bibnamefont{Weisbuch}}, \bibinfo
  {author} {\bibfnamefont{M.}~\bibnamefont{Nishioka}}, \bibinfo {author}
  {\bibfnamefont{A.}~\bibnamefont{Ishikawa}},\ and\ \bibinfo {author}
  {\bibfnamefont{Y.}~\bibnamefont{Arakawa}},\ }%
  \bibfield{title}{%
  \enquote{\bibinfo {title} {Observation of the coupled exciton-photon mode
  splitting in a semiconductor quantum microcavity},}\ }%
  \bibfield{journal}{%
  \Doi{10.1103/PhysRevLett.69.3314}{\bibinfo {journal} {Phys. Rev. Lett.}}\ }%
  \textbf{\bibinfo {volume} {69}},\ \bibinfo {pages} {3314--3317} (\bibinfo
  {year} {1992})%
  \bibAnnoteFile{NoStop}{Weisbuch_Polaritons}%
\bibitem{Kavokin_Microcavities}%
  \BibitemOpen
  \bibfield{author}{%
  \bibinfo {author} {\bibfnamefont{Alexey}\ \bibnamefont{Kavokin}}, \bibinfo
  {author} {\bibfnamefont{Jeremy~J.}\ \bibnamefont{Baumberg}}, \bibinfo
  {author} {\bibfnamefont{Guillaume}\ \bibnamefont{Malpuech}},\ and\ \bibinfo
  {author} {\bibfnamefont{Fabrice~P}\ \bibnamefont{Laussy}},\ }%
  \emph{\bibinfo {title} {Microcavities}}\ (\bibinfo {publisher} {Oxford
  University Press},\ \bibinfo {year} {2007})%
  \bibAnnoteFile{NoStop}{Kavokin_Microcavities}%
\bibitem{Kasprzak_BEC}%
  \BibitemOpen
  \bibfield{author}{%
  \bibinfo {author} {\bibfnamefont{J.}~\bibnamefont{Kasprzak}}, \bibinfo
  {author} {\bibfnamefont{M.}~\bibnamefont{Richard}}, \bibinfo {author}
  {\bibfnamefont{S.}~\bibnamefont{Kundermann}}, \bibinfo {author}
  {\bibfnamefont{A.}~\bibnamefont{Baas}}, \bibinfo {author}
  {\bibfnamefont{P.}~\bibnamefont{Jeambrun}}, \bibinfo {author}
  {\bibfnamefont{J.~M.~J.}\ \bibnamefont{Keeling}}, \bibinfo {author}
  {\bibfnamefont{F.~M.}\ \bibnamefont{Marchetti}}, \bibinfo {author}
  {\bibfnamefont{M.~H.}\ \bibnamefont{Szyma\'{n}ska}}, \bibinfo {author}
  {\bibfnamefont{R.}~\bibnamefont{Andr\'{e}}}, \bibinfo {author}
  {\bibfnamefont{J.~L.}\ \bibnamefont{Staehli}}, \bibinfo {author}
  {\bibfnamefont{V.}~\bibnamefont{Savona}}, \bibinfo {author}
  {\bibfnamefont{P.~B.}\ \bibnamefont{Littlewood}}, \bibinfo {author}
  {\bibfnamefont{B.}~\bibnamefont{Deveaud}},\ and\ \bibinfo {author}
  {\bibfnamefont{Le~Si}\ \bibnamefont{Dang}},\ }%
  \bibfield{title}{%
  \enquote{\bibinfo {title} {Bose–einstein condensation of exciton
  polaritons},}\ }%
  \bibfield{journal}{%
  \Doi{0.1038/nature05131}{\bibinfo {journal} {Nature}}\ }%
  \textbf{\bibinfo {volume} {443}},\ \bibinfo {pages} {409--414} (\bibinfo
  {year} {2006})%
  \bibAnnoteFile{NoStop}{Kasprzak_BEC}%
\bibitem{Yamamoto_PowerLawDecay}%
  \BibitemOpen
  \bibfield{author}{%
  \bibinfo {author} {\bibfnamefont{Georgios}\ \bibnamefont{Roumpos}}, \bibinfo
  {author} {\bibfnamefont{Michael}\ \bibnamefont{Lohse}}, \bibinfo {author}
  {\bibfnamefont{Wolfgang~H.}\ \bibnamefont{Nitsche}}, \bibinfo {author}
  {\bibfnamefont{Jonathan}\ \bibnamefont{Keeling}}, \bibinfo {author}
  {\bibfnamefont{Marzena~Hanna}\ \bibnamefont{Szyma{\'n}ska}}, \bibinfo
  {author} {\bibfnamefont{Peter~B.}\ \bibnamefont{Littlewood}}, \bibinfo
  {author} {\bibfnamefont{Andreas}\ \bibnamefont{L{\''o}ffler}}, \bibinfo
  {author} {\bibfnamefont{Sven}\ \bibnamefont{H{''o}fling}}, \bibinfo {author}
  {\bibfnamefont{Lukas}\ \bibnamefont{Worschech}}, \bibinfo {author}
  {\bibfnamefont{Alfred}\ \bibnamefont{Forchel}},\ and\ \bibinfo {author}
  {\bibfnamefont{Yoshihisa}\ \bibnamefont{Yamamoto}},\ }%
  \bibfield{title}{%
  \enquote{\bibinfo {title} {Power-law decay of the spatial correlation
  function in exciton-polariton condensates},}\ }%
  \bibfield{journal}{%
  \bibinfo {journal} {PNAS}\ }%
  \textbf{\bibinfo {volume} {109}},\ \bibinfo {pages} {6467--6472} (\bibinfo
  {year} {2012})%
  \bibAnnoteFile{NoStop}{Yamamoto_PowerLawDecay}%
\bibitem{Yamamoto_NPReview}%
  \BibitemOpen
  \bibfield{author}{%
  \bibinfo {author} {\bibfnamefont{Na~Young~Kim}\ \bibnamefont{Tim~Byrnes}}\
  and\ \bibinfo {author} {\bibfnamefont{Yoshihisa}\ \bibnamefont{Yamamoto}},\
  }%
  \bibfield{title}{%
  \enquote{\bibinfo {title} {Exciton–polariton condensates},}\ }%
  \bibfield{journal}{%
  \bibinfo {journal} {Nat. Phys.}\ }%
  \textbf{\bibinfo {volume} {10}},\ \bibinfo {pages} {803--813} (\bibinfo
  {year} {2014})%
  \bibAnnoteFile{NoStop}{Yamamoto_NPReview}%
\bibitem{Altman_DrivenSuperfluid2D}%
  \BibitemOpen
  \bibfield{author}{%
  \bibinfo {author} {\bibfnamefont{Ehud}\ \bibnamefont{Altman}}, \bibinfo
  {author} {\bibfnamefont{Lukas~M.}\ \bibnamefont{Sieberer}}, \bibinfo {author}
  {\bibfnamefont{Leiming}\ \bibnamefont{Chen}}, \bibinfo {author}
  {\bibfnamefont{Sebastian}\ \bibnamefont{Diehl}},\ and\ \bibinfo {author}
  {\bibfnamefont{John}\ \bibnamefont{Toner}},\ }%
  \bibfield{title}{%
  \enquote{\bibinfo {title} {Two-dimensional superfluidity of exciton
  polaritons requires strong anisotropy},}\ }%
  \bibfield{journal}{%
  \Doi{10.1103/PhysRevX.5.011017}{\bibinfo {journal} {Phys. Rev. X}}\ }%
  \textbf{\bibinfo {volume} {5}},\ \bibinfo {pages} {011017} (\bibinfo {year}
  {2015})%
  \bibAnnoteFile{NoStop}{Altman_DrivenSuperfluid2D}%
\bibitem{Grandjean_RoomTempLasing}%
  \BibitemOpen
  \bibfield{author}{%
  \bibinfo {author} {\bibfnamefont{S.}~\bibnamefont{Christopoulos}}, \bibinfo
  {author} {\bibfnamefont{G.~Baldassarri~H\"oger}\ \bibnamefont{von
  H\"ogersthal}}, \bibinfo {author} {\bibfnamefont{A.~J.~D.}\
  \bibnamefont{Grundy}}, \bibinfo {author} {\bibfnamefont{P.~G.}\
  \bibnamefont{Lagoudakis}}, \bibinfo {author} {\bibfnamefont{A.~V.}\
  \bibnamefont{Kavokin}}, \bibinfo {author} {\bibfnamefont{J.~J.}\
  \bibnamefont{Baumberg}}, \bibinfo {author}
  {\bibfnamefont{G.}~\bibnamefont{Christmann}}, \bibinfo {author}
  {\bibfnamefont{R.}~\bibnamefont{Butt\'e}}, \bibinfo {author}
  {\bibfnamefont{E.}~\bibnamefont{Feltin}}, \bibinfo {author}
  {\bibfnamefont{J.-F.}\ \bibnamefont{Carlin}},\ and\ \bibinfo {author}
  {\bibfnamefont{N.}~\bibnamefont{Grandjean}},\ }%
  \bibfield{title}{%
  \enquote{\bibinfo {title} {Room-temperature polariton lasing in semiconductor
  microcavities},}\ }%
  \bibfield{journal}{%
  \Doi{10.1103/PhysRevLett.98.126405}{\bibinfo {journal} {Phys. Rev. Lett.}}\
  }%
  \textbf{\bibinfo {volume} {98}},\ \bibinfo {pages} {126405} (\bibinfo {year}
  {2007})%
  \bibAnnoteFile{NoStop}{Grandjean_RoomTempLasing}%
\bibitem{Battacharya_RoomTemperatureElectrically}%
  \BibitemOpen
  \bibfield{author}{%
  \bibinfo {author} {\bibfnamefont{Pallab}\ \bibnamefont{Bhattacharya}},
  \bibinfo {author} {\bibfnamefont{Thomas}\ \bibnamefont{Frost}}, \bibinfo
  {author} {\bibfnamefont{Saniya}\ \bibnamefont{Deshpande}}, \bibinfo {author}
  {\bibfnamefont{Md~Zunaid}\ \bibnamefont{Baten}}, \bibinfo {author}
  {\bibfnamefont{Arnab}\ \bibnamefont{Hazari}},\ and\ \bibinfo {author}
  {\bibfnamefont{Ayan}\ \bibnamefont{Das}},\ }%
  \bibfield{title}{%
  \enquote{\bibinfo {title} {Room temperature electrically injected polariton
  laser},}\ }%
  \bibfield{journal}{%
  \Doi{10.1103/PhysRevLett.112.236802}{\bibinfo {journal} {Phys. Rev. Lett.}}\
  }%
  \textbf{\bibinfo {volume} {112}},\ \bibinfo {pages} {236802} (\bibinfo {year}
  {2014})%
  \bibAnnoteFile{NoStop}{Battacharya_RoomTemperatureElectrically}%
\bibitem{Bloch_interferometer}%
  \BibitemOpen
  \bibfield{author}{%
  \bibinfo {author} {\bibfnamefont{C.}~\bibnamefont{Sturm}}, \bibinfo {author}
  {\bibfnamefont{D.}~\bibnamefont{Tanese}}, \bibinfo {author}
  {\bibfnamefont{H.S.}\ \bibnamefont{Nguyen}}, \bibinfo {author}
  {\bibfnamefont{H.}~\bibnamefont{Flayac}}, \bibinfo {author}
  {\bibfnamefont{E.}~\bibnamefont{Galopin}}, \bibinfo {author}
  {\bibfnamefont{A.}~\bibnamefont{Lemaitre}}, \bibinfo {author}
  {\bibfnamefont{A.}~\bibnamefont{Amo}}, \bibinfo {author}
  {\bibfnamefont{G.}~\bibnamefont{Malpuech}},\ and\ \bibinfo {author}
  {\bibfnamefont{J.}~\bibnamefont{Bloch}},\ }%
  \bibfield{title}{%
  \enquote{\bibinfo {title} {All-optical phase modulation in a cavity-polariton
  mach–zehnder interferometer},}\ }%
  \bibfield{journal}{%
  \bibinfo {journal} {Nat. Commun.},\ \bibinfo {pages} {5:3278}}%
   (\bibinfo {year} {2014})%
  \bibAnnoteFile{NoStop}{Bloch_interferometer}%
\bibitem{Bramati_SpinSwitches}%
  \BibitemOpen
  \bibfield{author}{%
  \bibinfo {author} {\bibfnamefont{A.}~\bibnamefont{Amo}}, \bibinfo {author}
  {\bibfnamefont{T.~C.~H.}\ \bibnamefont{Liew}}, \bibinfo {author}
  {\bibfnamefont{C.}~\bibnamefont{Adrados}}, \bibinfo {author}
  {\bibfnamefont{R.}~\bibnamefont{Houdr{\'e}}}, \bibinfo {author}
  {\bibfnamefont{E.}~\bibnamefont{Giacobino}}, \bibinfo {author}
  {\bibfnamefont{A.~V.}\ \bibnamefont{Kavokin}},\ and\ \bibinfo {author}
  {\bibfnamefont{A.}~\bibnamefont{Bramati}},\ }%
  \bibfield{title}{%
  \enquote{\bibinfo {title} {Exciton–polariton spin switches},}\ }%
  \bibfield{journal}{%
  \Doi{10.1038/NPHOTON.2010.79}{\bibinfo {journal} {Nat. Photon.}}\ }%
  \textbf{\bibinfo {volume} {4}},\ \bibinfo {pages} {361--366} (\bibinfo {year}
  {2010})%
  \bibAnnoteFile{NoStop}{Bramati_SpinSwitches}%
\bibitem{Savvidis_TransistorSwitch}%
  \BibitemOpen
  \bibfield{author}{%
  \bibinfo {author} {\bibfnamefont{T.}~\bibnamefont{Gao}}, \bibinfo {author}
  {\bibfnamefont{P.~S.}\ \bibnamefont{Eldridge}}, \bibinfo {author}
  {\bibfnamefont{T.~C.~H.}\ \bibnamefont{Liew}}, \bibinfo {author}
  {\bibfnamefont{S.~I.}\ \bibnamefont{Tsintzos}}, \bibinfo {author}
  {\bibfnamefont{G.}~\bibnamefont{Stavrinidis}}, \bibinfo {author}
  {\bibfnamefont{G.}~\bibnamefont{Deligeorgis}}, \bibinfo {author}
  {\bibfnamefont{Z.}~\bibnamefont{Hatzopoulos}},\ and\ \bibinfo {author}
  {\bibfnamefont{P.~G.}\ \bibnamefont{Savvidis}},\ }%
  \bibfield{title}{%
  \enquote{\bibinfo {title} {Polariton condensate transistor switch},}\ }%
  \bibfield{journal}{%
  \Doi{10.1103/PhysRevB.85.235102}{\bibinfo {journal} {Phys. Rev. B}}\ }%
  \textbf{\bibinfo {volume} {85}},\ \bibinfo {pages} {235102} (\bibinfo {year}
  {2012})%
  \bibAnnoteFile{NoStop}{Savvidis_TransistorSwitch}%
\bibitem{Sanvitto_Transistor}%
  \BibitemOpen
  \bibfield{author}{%
  \bibinfo {author} {\bibfnamefont{D.}~\bibnamefont{Ballarini}}, \bibinfo
  {author} {\bibfnamefont{M.~De}\ \bibnamefont{Giorgi}}, \bibinfo {author}
  {\bibfnamefont{E.}~\bibnamefont{Cancellieri}}, \bibinfo {author}
  {\bibfnamefont{R.}~\bibnamefont{Houdr\'{e}}}, \bibinfo {author}
  {\bibfnamefont{E.}~\bibnamefont{Giacobino}}, \bibinfo {author}
  {\bibfnamefont{R.}~\bibnamefont{Cingolani}}, \bibinfo {author}
  {\bibfnamefont{A.}~\bibnamefont{Bramati}}, \bibinfo {author}
  {\bibfnamefont{G.}~\bibnamefont{Gigli}},\ and\ \bibinfo {author}
  {\bibfnamefont{D.}~\bibnamefont{Sanvitto}},\ }%
  \bibfield{title}{%
  \enquote{\bibinfo {title} {All-optical polariton transistor},}\ }%
  \bibfield{journal}{%
  \Doi{10.1038/ncomms2734}{\bibinfo {journal} {Nat. Commun.}}\ }%
  \textbf{\bibinfo {volume} {4}},\ \bibinfo {pages} {1778} (\bibinfo {year}
  {2013})%
  \bibAnnoteFile{NoStop}{Sanvitto_Transistor}%
\bibitem{Amo_Superfluidity}%
  \BibitemOpen
  \bibfield{author}{%
  \bibinfo {author} {\bibfnamefont{Alberto}\ \bibnamefont{Amo}}, \bibinfo
  {author} {\bibfnamefont{J{\'e}r{\^o}me}\ \bibnamefont{Lefr{\'e}re}}, \bibinfo
  {author} {\bibfnamefont{Simon}\ \bibnamefont{Pigeon}}, \bibinfo {author}
  {\bibfnamefont{Claire}\ \bibnamefont{Adrados}}, \bibinfo {author}
  {\bibfnamefont{Cristiano}\ \bibnamefont{Ciuti}}, \bibinfo {author}
  {\bibfnamefont{Iacopo}\ \bibnamefont{Carusotto}}, \bibinfo {author}
  {\bibfnamefont{Romuald}\ \bibnamefont{Houdr{\'e}}}, \bibinfo {author}
  {\bibfnamefont{Elisabeth}\ \bibnamefont{Giacobino}},\ and\ \bibinfo {author}
  {\bibfnamefont{Alberto}\ \bibnamefont{Bramati}},\ }%
  \bibfield{title}{%
  \enquote{\bibinfo {title} {Superfluidity of polaritons in semiconductor
  microcavities},}\ }%
  \bibfield{journal}{%
  \Doi{10.1038/nphys1364}{\bibinfo {journal} {Nature Physics}}\ }%
  \textbf{\bibinfo {volume} {5}},\ \bibinfo {pages} {805 -- 810} (\bibinfo
  {year} {2009})%
  \bibAnnoteFile{NoStop}{Amo_Superfluidity}%
\bibitem{Deveaud_Vortices}%
  \BibitemOpen
  \bibfield{author}{%
  \bibinfo {author} {\bibfnamefont{K.~G.}\ \bibnamefont{Lagoudakis}}, \bibinfo
  {author} {\bibfnamefont{M.}~\bibnamefont{Wouters}}, \bibinfo {author}
  {\bibfnamefont{M.}~\bibnamefont{Richard}}, \bibinfo {author}
  {\bibfnamefont{A.}~\bibnamefont{Baas}}, \bibinfo {author}
  {\bibfnamefont{I.}~\bibnamefont{Carusotto}}, \bibinfo {author}
  {\bibfnamefont{R.}~\bibnamefont{Andr\'e}}, \bibinfo {author}
  {\bibfnamefont{Le~Si}\ \bibnamefont{Dang}},\ and\ \bibinfo {author}
  {\bibfnamefont{B.}~\bibnamefont{Deveaud-Pl\'edran}},\ }%
  \bibfield{title}{%
  \enquote{\bibinfo {title} {Quantized vortices in an exciton–polariton
  condensate},}\ }%
  \bibfield{journal}{%
  \Doi{10.1038/nphys1051}{\bibinfo {journal} {Nat. Phys.}}\ }%
  \textbf{\bibinfo {volume} {4}},\ \bibinfo {pages} {706--710} (\bibinfo {year}
  {2008})%
  \bibAnnoteFile{NoStop}{Deveaud_Vortices}%
\bibitem{Nguyen_AcousticBlackHole}%
  \BibitemOpen
  \bibfield{author}{%
  \bibinfo {author} {\bibfnamefont{H.~S.}\ \bibnamefont{Nguyen}}, \bibinfo
  {author} {\bibfnamefont{D.}~\bibnamefont{Gerace}}, \bibinfo {author}
  {\bibfnamefont{I.}~\bibnamefont{Carusotto}}, \bibinfo {author}
  {\bibfnamefont{D.}~\bibnamefont{Sanvitto}}, \bibinfo {author}
  {\bibfnamefont{E.}~\bibnamefont{Galopin}}, \bibinfo {author}
  {\bibfnamefont{A.}~\bibnamefont{Lema\^{\i}tre}}, \bibinfo {author}
  {\bibfnamefont{I.}~\bibnamefont{Sagnes}}, \bibinfo {author}
  {\bibfnamefont{J.}~\bibnamefont{Bloch}},\ and\ \bibinfo {author}
  {\bibfnamefont{A.}~\bibnamefont{Amo}},\ }%
  \bibfield{title}{%
  \enquote{\bibinfo {title} {Acoustic black hole in a stationary hydrodynamic
  flow of microcavity polaritons},}\ }%
  \bibfield{journal}{%
  \Doi{10.1103/PhysRevLett.114.036402}{\bibinfo {journal} {Phys. Rev. Lett.}}\
  }%
  \textbf{\bibinfo {volume} {114}},\ \bibinfo {pages} {036402} (\bibinfo {year}
  {2015})%
  \bibAnnoteFile{NoStop}{Nguyen_AcousticBlackHole}%
\bibitem{Amo_HydrodynamicSolitons}%
  \BibitemOpen
  \bibfield{author}{%
  \bibinfo {author} {\bibfnamefont{A.}~\bibnamefont{Amo}}, \bibinfo {author}
  {\bibfnamefont{S.}~\bibnamefont{Pigeon}}, \bibinfo {author}
  {\bibfnamefont{D.}~\bibnamefont{Sanvitto}}, \bibinfo {author}
  {\bibfnamefont{V.~G.}\ \bibnamefont{Sala}}, \bibinfo {author}
  {\bibfnamefont{R.}~\bibnamefont{Hivet}}, \bibinfo {author}
  {\bibfnamefont{I.}~\bibnamefont{Carusotto}}, \bibinfo {author}
  {\bibfnamefont{F.}~\bibnamefont{Pisanello}}, \bibinfo {author}
  {\bibfnamefont{G.}~\bibnamefont{Lem{\'e}nager}}, \bibinfo {author}
  {\bibfnamefont{R.}~\bibnamefont{Houdr{\'e}}}, \bibinfo {author}
  {\bibfnamefont{E}~\bibnamefont{Giacobino}}, \bibinfo {author}
  {\bibfnamefont{C.}~\bibnamefont{Ciuti}},\ and\ \bibinfo {author}
  {\bibfnamefont{A.}~\bibnamefont{Bramati}},\ }%
  \bibfield{title}{%
  \enquote{\bibinfo {title} {Polariton superfluids reveal quantum hydrodynamic
  solitons},}\ }%
  \bibfield{journal}{%
  \Doi{10.1126/science.1202307}{\bibinfo {journal} {Science}}\ }%
  \textbf{\bibinfo {volume} {332}},\ \bibinfo {pages} {1167--1170} (\bibinfo
  {year} {2011}),\ ISSN \bibinfo {issn} {0036-8075}%
  \bibAnnoteFile{NoStop}{Amo_HydrodynamicSolitons}%
\bibitem{Santos_BrightSoliton}%
  \BibitemOpen
  \bibfield{author}{%
  \bibinfo {author} {\bibfnamefont{M.}~\bibnamefont{Sich}}, \bibinfo {author}
  {\bibfnamefont{D.~N.}\ \bibnamefont{Krizhanovskii}}, \bibinfo {author}
  {\bibfnamefont{M.~S.}\ \bibnamefont{Skolnick}}, \bibinfo {author}
  {\bibfnamefont{A.~V.}\ \bibnamefont{Gorbach}}, \bibinfo {author}
  {\bibfnamefont{R.}~\bibnamefont{Hartley}}, \bibinfo {author}
  {\bibfnamefont{D.~V.}\ \bibnamefont{Skryabin}}, \bibinfo {author}
  {\bibfnamefont{E.~A.}\ \bibnamefont{Cerda-M\'endez}}, \bibinfo {author}
  {\bibfnamefont{K.}~\bibnamefont{Biermann}}, \bibinfo {author}
  {\bibfnamefont{R.}~\bibnamefont{Hey}},\ and\ \bibinfo {author}
  {\bibfnamefont{P.~V.}\ \bibnamefont{Santos}},\ }%
  \bibfield{title}{%
  \enquote{\bibinfo {title} {Observation of bright polariton solitons in a
  semiconductor microcavity},}\ }%
  \bibfield{journal}{%
  \bibinfo {journal} {Nature Photon.}\ }%
  \textbf{\bibinfo {volume} {6}},\ \bibinfo {pages} {50--55} (\bibinfo {year}
  {2012})%
  \bibAnnoteFile{NoStop}{Santos_BrightSoliton}%
\bibitem{Kena_Organic}%
  \BibitemOpen
  \bibfield{author}{%
  \bibinfo {author} {\bibfnamefont{S.}~\bibnamefont{K\'{e}na-Cohen}}\ and\
  \bibinfo {author} {\bibfnamefont{S.~R.}\ \bibnamefont{Forrest}},\ }%
  \bibfield{title}{%
  \enquote{\bibinfo {title} {Room-temperature polariton lasing in an organic
  single-crystal microcavity},}\ }%
  \bibfield{journal}{%
  \Doi{10.1038/NPHOTON.2010.86}{\bibinfo {journal} {Nat. Photon.}}\ }%
  \textbf{\bibinfo {volume} {4}},\ \bibinfo {pages} {371--375} (\bibinfo {year}
  {2010})%
  \bibAnnoteFile{NoStop}{Kena_Organic}%
\bibitem{Mahrt_RTCondensatePolymer}%
  \BibitemOpen
  \bibfield{author}{%
  \bibinfo {author} {\bibfnamefont{Johannes~D.}\ \bibnamefont{Plumhof}},
  \bibinfo {author} {\bibfnamefont{Thilo}\ \bibnamefont{St\"{o}ferle}},
  \bibinfo {author} {\bibfnamefont{Lijian}\ \bibnamefont{Mai}}, \bibinfo
  {author} {\bibfnamefont{Ullrich}\ \bibnamefont{Scherf}},\ and\ \bibinfo
  {author} {\bibfnamefont{Rainer~F.}\ \bibnamefont{Mahrt}},\ }%
  \bibfield{title}{%
  \enquote{\bibinfo {title} {Room-temperature bose–einstein condensation of
  cavity exciton–polaritons in a polymer},}\ }%
  \bibfield{journal}{%
  \Doi{10.1038/NMAT3825}{\bibinfo {journal} {Nat. Mater.}}\ }%
  \textbf{\bibinfo {volume} {13}},\ \bibinfo {pages} {247--252} (\bibinfo
  {year} {2014})%
  \bibAnnoteFile{NoStop}{Mahrt_RTCondensatePolymer}%
\bibitem{KenaCohen_NonlinearOrganic}%
  \BibitemOpen
  \bibfield{author}{%
  \bibinfo {author} {\bibfnamefont{K.~S.}\ \bibnamefont{Daskalakis}}, \bibinfo
  {author} {\bibfnamefont{S.~A.}\ \bibnamefont{Maier}},\ and\ \bibinfo {author}
  {\bibfnamefont{R.~Murrayand~S.}\ \bibnamefont{Kena-Cohen}},\ }%
  \bibfield{title}{%
  \enquote{\bibinfo {title} {Nonlinear interactions in an organic polariton
  condensate},}\ }%
  \bibfield{journal}{%
  \bibinfo {journal} {Nat. Mater.}\ }%
  \textbf{\bibinfo {volume} {13}},\ \bibinfo {pages} {271--278} (\bibinfo
  {year} {2014})%
  \bibAnnoteFile{NoStop}{KenaCohen_NonlinearOrganic}%
\bibitem{Hofling_ElectricallyPumped}%
  \BibitemOpen
  \bibfield{author}{%
  \bibinfo {author} {\bibfnamefont{C.}~\bibnamefont{Schneider}}, \bibinfo
  {author} {\bibfnamefont{A.}~\bibnamefont{Rahimi-Iman}}, \bibinfo {author}
  {\bibfnamefont{NY.}\ \bibnamefont{Kim}}, \bibinfo {author}
  {\bibfnamefont{J.}~\bibnamefont{Fischer}}, \bibinfo {author}
  {\bibfnamefont{IG.}\ \bibnamefont{Savenko}}, \bibinfo {author}
  {\bibfnamefont{M.}~\bibnamefont{Amthor}}, \bibinfo {author}
  {\bibfnamefont{M.}~\bibnamefont{Lermer}}, \bibinfo {author}
  {\bibfnamefont{A.}~\bibnamefont{Wolf}}, \bibinfo {author}
  {\bibfnamefont{L.}~\bibnamefont{Worschech}}, \bibinfo {author}
  {\bibfnamefont{VD.}\ \bibnamefont{Kulakovskii}}, \bibinfo {author}
  {\bibfnamefont{IA.}\ \bibnamefont{Shelykh}}, \bibinfo {author}
  {\bibfnamefont{M.}~\bibnamefont{Kamp}}, \bibinfo {author}
  {\bibfnamefont{S.}~\bibnamefont{Reitzenstein}}, \bibinfo {author}
  {\bibfnamefont{A.}~\bibnamefont{Forchel}}, \bibinfo {author}
  {\bibfnamefont{Y.}~\bibnamefont{Yamamoto}},\ and\ \bibinfo {author}
  {\bibfnamefont{S.}~\bibnamefont{H{\"o}fling}},\ }%
  \bibfield{title}{%
  \enquote{\bibinfo {title} {An electrically pumped polariton laser},}\ }%
  \bibfield{journal}{%
  \Doi{10.1038/nature12036}{\bibinfo {journal} {Nature}}\ }%
  \textbf{\bibinfo {volume} {497}},\ \bibinfo {pages} {348--352} (\bibinfo
  {year} {2013})%
  \bibAnnoteFile{NoStop}{Hofling_ElectricallyPumped}%
\bibitem{Haug_QuantumKineticGPDerivation}%
  \BibitemOpen
  \bibfield{author}{%
  \bibinfo {author} {\bibfnamefont{H.}~\bibnamefont{Haug}}, \bibinfo {author}
  {\bibfnamefont{T.~D.}\ \bibnamefont{Doan}},\ and\ \bibinfo {author}
  {\bibfnamefont{D.~B.}\ \bibnamefont{Tran~Thoai}},\ }%
  \bibfield{title}{%
  \enquote{\bibinfo {title} {Quantum kinetic derivation of the nonequilibrium
  gross-pitaevskii equation for nonresonant excitation of microcavity
  polaritons},}\ }%
  \bibfield{journal}{%
  \Doi{10.1103/PhysRevB.89.155302}{\bibinfo {journal} {Phys. Rev. B}}\ }%
  \textbf{\bibinfo {volume} {89}},\ \bibinfo {pages} {155302} (\bibinfo {year}
  {2014})%
  \bibAnnoteFile{NoStop}{Haug_QuantumKineticGPDerivation}%
\bibitem{Wouters_ClassicalFields}%
  \BibitemOpen
  \bibfield{author}{%
  \bibinfo {author} {\bibfnamefont{Michiel}\ \bibnamefont{Wouters}}\ and\
  \bibinfo {author} {\bibfnamefont{Vincenzo}\ \bibnamefont{Savona}},\ }%
  \bibfield{title}{%
  \enquote{\bibinfo {title} {Stochastic classical field model for polariton
  condensates},}\ }%
  \bibfield{journal}{%
  \Doi{10.1103/PhysRevB.79.165302}{\bibinfo {journal} {Phys. Rev. B}}\ }%
  \textbf{\bibinfo {volume} {79}},\ \bibinfo {pages} {165302} (\bibinfo {year}
  {2009})%
  \bibAnnoteFile{NoStop}{Wouters_ClassicalFields}%
\bibitem{Malpuech_Hybrid}%
  \BibitemOpen
  \bibfield{author}{%
  \bibinfo {author} {\bibfnamefont{D.~D.}\ \bibnamefont{Solnyshkov}}, \bibinfo
  {author} {\bibfnamefont{H.}~\bibnamefont{Tercas}}, \bibinfo {author}
  {\bibfnamefont{K.}~\bibnamefont{Dini}},\ and\ \bibinfo {author}
  {\bibfnamefont{G.}~\bibnamefont{Malpuech}},\ }%
  \bibfield{title}{%
  \enquote{\bibinfo {title} {Hybrid boltzmann-gross-pitaevskii theory of
  bose-einstein condensation and superfluidity in open driven-dissipative
  systems},}\ }%
  \bibfield{journal}{%
  \Doi{10.1103/PhysRevA.89.033626}{\bibinfo {journal} {Phys. Rev. A}}\ }%
  \textbf{\bibinfo {volume} {89}},\ \bibinfo {pages} {033626} (\bibinfo {year}
  {2014})%
  \bibAnnoteFile{NoStop}{Malpuech_Hybrid}%
\bibitem{Laussy_SpontaneousCoherence}%
  \BibitemOpen
  \bibfield{author}{%
  \bibinfo {author} {\bibfnamefont{Fabrice~P.}\ \bibnamefont{Laussy}}, \bibinfo
  {author} {\bibfnamefont{G.}~\bibnamefont{Malpuech}}, \bibinfo {author}
  {\bibfnamefont{A.}~\bibnamefont{Kavokin}},\ and\ \bibinfo {author}
  {\bibfnamefont{P.}~\bibnamefont{Bigenwald}},\ }%
  \bibfield{title}{%
  \enquote{\bibinfo {title} {Spontaneous coherence buildup in a polariton
  laser},}\ }%
  \bibfield{journal}{%
  \Doi{10.1103/PhysRevLett.93.016402}{\bibinfo {journal} {Phys. Rev. Lett.}}\
  }%
  \textbf{\bibinfo {volume} {93}},\ \bibinfo {pages} {016402} (\bibinfo {year}
  {2004})%
  \bibAnnoteFile{NoStop}{Laussy_SpontaneousCoherence}%
\bibitem{Bloch_Molecules}%
  \BibitemOpen
  \bibfield{author}{%
  \bibinfo {author} {\bibfnamefont{Marta}\ \bibnamefont{Galbiati}}, \bibinfo
  {author} {\bibfnamefont{Lydie}\ \bibnamefont{Ferrier}}, \bibinfo {author}
  {\bibfnamefont{Dmitry~D.}\ \bibnamefont{Solnyshkov}}, \bibinfo {author}
  {\bibfnamefont{Dimitrii}\ \bibnamefont{Tanese}}, \bibinfo {author}
  {\bibfnamefont{Esther}\ \bibnamefont{Wertz}}, \bibinfo {author}
  {\bibfnamefont{Alberto}\ \bibnamefont{Amo}}, \bibinfo {author}
  {\bibfnamefont{Marco}\ \bibnamefont{Abbarchi}}, \bibinfo {author}
  {\bibfnamefont{Pascale}\ \bibnamefont{Senellart}}, \bibinfo {author}
  {\bibfnamefont{Isabelle}\ \bibnamefont{Sagnes}}, \bibinfo {author}
  {\bibfnamefont{Aristide}\ \bibnamefont{Lema\^itre}}, \bibinfo {author}
  {\bibfnamefont{Elisabeth}\ \bibnamefont{Galopin}}, \bibinfo {author}
  {\bibfnamefont{Guillaume}\ \bibnamefont{Malpuech}},\ and\ \bibinfo {author}
  {\bibfnamefont{Jacqueline}\ \bibnamefont{Bloch}},\ }%
  \bibfield{title}{%
  \enquote{\bibinfo {title} {Polariton condensation in photonic molecules},}\
  }%
  \bibfield{journal}{%
  \Doi{10.1103/PhysRevLett.108.126403}{\bibinfo {journal} {Phys. Rev. Lett.}}\
  }%
  \textbf{\bibinfo {volume} {108}},\ \bibinfo {pages} {126403} (\bibinfo {year}
  {2012})%
  \bibAnnoteFile{NoStop}{Bloch_Molecules}%
\bibitem{Tassone_Bottleneck}%
  \BibitemOpen
  \bibfield{author}{%
  \bibinfo {author} {\bibfnamefont{F.}~\bibnamefont{Tassone}}, \bibinfo
  {author} {\bibfnamefont{C.}~\bibnamefont{Piermarocchi}}, \bibinfo {author}
  {\bibfnamefont{V.}~\bibnamefont{Savona}}, \bibinfo {author}
  {\bibfnamefont{A.}~\bibnamefont{Quattropani}},\ and\ \bibinfo {author}
  {\bibfnamefont{P.}~\bibnamefont{Schwendimann}},\ }%
  \bibfield{title}{%
  \enquote{\bibinfo {title} {Bottleneck effects in the relaxation and
  photoluminescence of microcavity polaritons},}\ }%
  \bibfield{journal}{%
  \Doi{10.1103/PhysRevB.56.7554}{\bibinfo {journal} {Phys. Rev. B}}\ }%
  \textbf{\bibinfo {volume} {56}},\ \bibinfo {pages} {7554--7563} (\bibinfo
  {year} {1997})%
  \bibAnnoteFile{NoStop}{Tassone_Bottleneck}%
\bibitem{Wouters_ExcitationSpectrum}%
  \BibitemOpen
  \bibfield{author}{%
  \bibinfo {author} {\bibfnamefont{Michiel}\ \bibnamefont{Wouters}}\ and\
  \bibinfo {author} {\bibfnamefont{Iacopo}\ \bibnamefont{Carusotto}},\ }%
  \bibfield{title}{%
  \enquote{\bibinfo {title} {Excitations in a nonequilibrium bose-einstein
  condensate of exciton polaritons},}\ }%
  \bibfield{journal}{%
  \Doi{10.1103/PhysRevLett.99.140402}{\bibinfo {journal} {Phys. Rev. Lett.}}\
  }%
  \textbf{\bibinfo {volume} {99}},\ \bibinfo {pages} {140402} (\bibinfo {year}
  {2007})%
  \bibAnnoteFile{NoStop}{Wouters_ExcitationSpectrum}%
\bibitem{Kneer_GenericAtomLaser}%
  \BibitemOpen
  \bibfield{author}{%
  \bibinfo {author} {\bibfnamefont{B.}~\bibnamefont{Kneer}}, \bibinfo {author}
  {\bibfnamefont{T.}~\bibnamefont{Wong}}, \bibinfo {author}
  {\bibfnamefont{K.}~\bibnamefont{Vogel}}, \bibinfo {author}
  {\bibfnamefont{W.~P}\ \bibnamefont{Schleich}},\ and\ \bibinfo {author}
  {\bibfnamefont{D.~F.}\ \bibnamefont{Walls}},\ }%
  \bibfield{title}{%
  \enquote{\bibinfo {title} {Generic model of an atom laser},}\ }%
  \bibfield{journal}{%
  \bibinfo {journal} {Phys. Rev. A}\ }%
  \textbf{\bibinfo {volume} {58}},\ \bibinfo {pages} {4841 -- 4853} (\bibinfo
  {year} {1998})%
  \bibAnnoteFile{NoStop}{Kneer_GenericAtomLaser}%
\bibitem{Wouters_EnergyRelaxation}%
  \BibitemOpen
  \bibfield{author}{%
  \bibinfo {author} {\bibfnamefont{M.}~\bibnamefont{Wouters}}, \bibinfo
  {author} {\bibfnamefont{T.~C.~H.}\ \bibnamefont{Liew}},\ and\ \bibinfo
  {author} {\bibfnamefont{V.}~\bibnamefont{Savona}},\ }%
  \bibfield{title}{%
  \enquote{\bibinfo {title} {Energy relaxation in one-dimensional polariton
  condensates},}\ }%
  \bibfield{journal}{%
  \Doi{10.1103/PhysRevB.82.245315}{\bibinfo {journal} {Phys. Rev. B}}\ }%
  \textbf{\bibinfo {volume} {82}},\ \bibinfo {pages} {245315} (\bibinfo {year}
  {2010})%
  \bibAnnoteFile{NoStop}{Wouters_EnergyRelaxation}%
\bibitem{Carusotto_NonequilibriumQuasicondensates}%
  \BibitemOpen
  \bibfield{author}{%
  \bibinfo {author} {\bibfnamefont{A.}~\bibnamefont{Chiocchetta}}\ and\
  \bibinfo {author} {\bibfnamefont{I.}~\bibnamefont{Carusotto}},\ }%
  \bibfield{title}{%
  \enquote{\bibinfo {title} {Non-equilibrium quasi-condensates in reduced
  dimensions},}\ }%
  \bibfield{journal}{%
  \bibinfo {journal} {EPL}\ }%
  \textbf{\bibinfo {volume} {102}},\ \bibinfo {pages} {67007} (\bibinfo {year}
  {2013})%
  \bibAnnoteFile{NoStop}{Carusotto_NonequilibriumQuasicondensates}%
\bibitem{Keeling_VortexDynamics}%
  \BibitemOpen
  \bibfield{author}{%
  \bibinfo {author} {\bibfnamefont{Jonathan}\ \bibnamefont{Keeling}}\ and\
  \bibinfo {author} {\bibfnamefont{Natalia~G.}\ \bibnamefont{Berloff}},\ }%
  \bibfield{title}{%
  \enquote{\bibinfo {title} {Spontaneous rotating vortex lattices in a pumped
  decaying condensate},}\ }%
  \bibfield{journal}{%
  \Doi{10.1103/PhysRevLett.100.250401}{\bibinfo {journal} {Phys. Rev. Lett.}}\
  }%
  \textbf{\bibinfo {volume} {100}},\ \bibinfo {pages} {250401} (\bibinfo {year}
  {2008})%
  \bibAnnoteFile{NoStop}{Keeling_VortexDynamics}%
\bibitem{Sieberer_DynamicalCritical}%
  \BibitemOpen
  \bibfield{author}{%
  \bibinfo {author} {\bibfnamefont{L.~M.}\ \bibnamefont{Sieberer}}, \bibinfo
  {author} {\bibfnamefont{S.~D.}\ \bibnamefont{Huber}}, \bibinfo {author}
  {\bibfnamefont{E.}~\bibnamefont{Altman}},\ and\ \bibinfo {author}
  {\bibfnamefont{S.}~\bibnamefont{Diehl}},\ }%
  \bibfield{title}{%
  \enquote{\bibinfo {title} {Dynamical critical phenomena in driven-dissipative
  systems},}\ }%
  \bibfield{journal}{%
  \Doi{10.1103/PhysRevLett.110.195301}{\bibinfo {journal} {Phys. Rev. Lett.}}\
  }%
  \textbf{\bibinfo {volume} {110}},\ \bibinfo {pages} {195301} (\bibinfo {year}
  {2013})%
  \bibAnnoteFile{NoStop}{Sieberer_DynamicalCritical}%
\bibitem{Bobrovska_Adiabatic}%
  \BibitemOpen
  \bibfield{author}{%
  \bibinfo {author} {\bibfnamefont{Nataliya}\ \bibnamefont{Bobrovska}}\ and\
  \bibinfo {author} {\bibfnamefont{Micha\l{}}\ \bibnamefont{Matuszewski}},\ }%
  \bibfield{title}{%
  \enquote{\bibinfo {title} {Adiabatic approximation and fluctuations in
  exciton-polariton condensates},}\ }%
  \bibfield{journal}{%
  \Doi{10.1103/PhysRevB.92.035311}{\bibinfo {journal} {Phys. Rev. B}}\ }%
  \textbf{\bibinfo {volume} {92}},\ \bibinfo {pages} {035311} (\bibinfo {year}
  {2015})%
  \bibAnnoteFile{NoStop}{Bobrovska_Adiabatic}%
\bibitem{Kena_SpatialCoherence}%
  \BibitemOpen
  \bibfield{author}{%
  \bibinfo {author} {\bibfnamefont{K.~S.}\ \bibnamefont{Daskalakis}}, \bibinfo
  {author} {\bibfnamefont{S.~A.}\ \bibnamefont{Maier}},\ and\ \bibinfo {author}
  {\bibfnamefont{S.}~\bibnamefont{K\'ena-Cohen}},\ }%
  \bibfield{title}{%
  \enquote{\bibinfo {title} {Spatial coherence and stability in a disordered
  organic polariton condensate},}\ }%
  \bibfield{journal}{%
  \Doi{10.1103/PhysRevLett.115.035301}{\bibinfo {journal} {Phys. Rev. Lett.}}\
  }%
  \textbf{\bibinfo {volume} {115}},\ \bibinfo {pages} {035301} (\bibinfo {year}
  {2015})%
  \bibAnnoteFile{NoStop}{Kena_SpatialCoherence}%
\bibitem{Ostrovskaya_DarkSoliton}%
  \BibitemOpen
  \bibfield{author}{%
  \bibinfo {author} {\bibfnamefont{Lev~A.}\ \bibnamefont{Smirnov}}, \bibinfo
  {author} {\bibfnamefont{Daria~A.}\ \bibnamefont{Smirnova}}, \bibinfo {author}
  {\bibfnamefont{Elena~A.}\ \bibnamefont{Ostrovskaya}},\ and\ \bibinfo {author}
  {\bibfnamefont{Yuri~S.}\ \bibnamefont{Kivshar}},\ }%
  \bibfield{title}{%
  \enquote{\bibinfo {title} {Dynamics and stability of dark solitons in
  exciton-polariton condensates},}\ }%
  \bibfield{journal}{%
  \Doi{10.1103/PhysRevB.89.235310}{\bibinfo {journal} {Phys. Rev. B}}\ }%
  \textbf{\bibinfo {volume} {89}},\ \bibinfo {pages} {235310} (\bibinfo {year}
  {2014})%
  \bibAnnoteFile{NoStop}{Ostrovskaya_DarkSoliton}%
\bibitem{Wouters_SpatialCoherence}%
  \BibitemOpen
  \bibfield{author}{%
  \bibinfo {author} {\bibfnamefont{Vladimir~N.}\ \bibnamefont{Gladilin}},
  \bibinfo {author} {\bibfnamefont{Kai}\ \bibnamefont{Ji}},\ and\ \bibinfo
  {author} {\bibfnamefont{Michiel}\ \bibnamefont{Wouters}},\ }%
  \bibfield{title}{%
  \enquote{\bibinfo {title} {Spatial coherence of weakly interacting
  one-dimensional nonequilibrium bosonic quantum fluids},}\ }%
  \bibfield{journal}{%
  \Doi{10.1103/PhysRevA.90.023615}{\bibinfo {journal} {Phys. Rev. A}}\ }%
  \textbf{\bibinfo {volume} {90}},\ \bibinfo {pages} {023615} (\bibinfo {year}
  {2014})%
  \bibAnnoteFile{NoStop}{Wouters_SpatialCoherence}%
\bibitem{Liew_InstabilityInduced}%
  \BibitemOpen
  \bibfield{author}{%
  \bibinfo {author} {\bibfnamefont{T.~C.~H.}\ \bibnamefont{Liew}}, \bibinfo
  {author} {\bibfnamefont{O.~A.}\ \bibnamefont{Egorov}}, \bibinfo {author}
  {\bibfnamefont{M.}~\bibnamefont{Matuszewski}}, \bibinfo {author}
  {\bibfnamefont{O.}~\bibnamefont{Kyriienko}}, \bibinfo {author}
  {\bibfnamefont{X.}~\bibnamefont{Ma}},\ and\ \bibinfo {author}
  {\bibfnamefont{E.~A.}\ \bibnamefont{Ostrovskaya}},\ }%
  \bibfield{title}{%
  \enquote{\bibinfo {title} {Instability-induced formation and nonequilibrium
  dynamics of phase defects in polariton condensates},}\ }%
  \bibfield{journal}{%
  \Doi{10.1103/PhysRevB.91.085413}{\bibinfo {journal} {Phys. Rev. B}}\ }%
  \textbf{\bibinfo {volume} {91}},\ \bibinfo {pages} {085413} (\bibinfo {year}
  {2015})%
  \bibAnnoteFile{NoStop}{Liew_InstabilityInduced}%
\bibitem{Byrnes_Yamamoto_NegativeBogoliubov}%
  \BibitemOpen
  \bibfield{author}{%
  \bibinfo {author} {\bibfnamefont{Tim}\ \bibnamefont{Byrnes}}, \bibinfo
  {author} {\bibfnamefont{Tomoyuki}\ \bibnamefont{Horikiri}}, \bibinfo {author}
  {\bibfnamefont{Natsuko}\ \bibnamefont{Ishida}}, \bibinfo {author}
  {\bibfnamefont{Michael}\ \bibnamefont{Fraser}},\ and\ \bibinfo {author}
  {\bibfnamefont{Yoshihisa}\ \bibnamefont{Yamamoto}},\ }%
  \bibfield{title}{%
  \enquote{\bibinfo {title} {Negative bogoliubov dispersion in
  exciton-polariton condensates},}\ }%
  \bibfield{journal}{%
  \Doi{10.1103/PhysRevB.85.075130}{\bibinfo {journal} {Phys. Rev. B}}\ }%
  \textbf{\bibinfo {volume} {85}},\ \bibinfo {pages} {075130} (\bibinfo {year}
  {2012})%
  \bibAnnoteFile{NoStop}{Byrnes_Yamamoto_NegativeBogoliubov}%
\bibitem{Bobrovska_Stability}%
  \BibitemOpen
  \bibfield{author}{%
  \bibinfo {author} {\bibfnamefont{Nataliya}\ \bibnamefont{Bobrovska}},
  \bibinfo {author} {\bibfnamefont{Elena~A.}\ \bibnamefont{Ostrovskaya}},\ and\
  \bibinfo {author} {\bibfnamefont{Micha\l{}}\ \bibnamefont{Matuszewski}},\ }%
  \bibfield{title}{%
  \enquote{\bibinfo {title} {Stability and spatial coherence of nonresonantly
  pumped exciton-polariton condensates},}\ }%
  \bibfield{journal}{%
  \Doi{10.1103/PhysRevB.90.205304}{\bibinfo {journal} {Phys. Rev. B}}\ }%
  \textbf{\bibinfo {volume} {90}},\ \bibinfo {pages} {205304} (\bibinfo {year}
  {2014})%
  \bibAnnoteFile{NoStop}{Bobrovska_Stability}%
\bibitem{Bloch_DynamicsElectronGas}%
  \BibitemOpen
  \bibfield{author}{%
  \bibinfo {author} {\bibfnamefont{D.}~\bibnamefont{Bajoni}}, \bibinfo {author}
  {\bibfnamefont{M.}~\bibnamefont{Perrin}}, \bibinfo {author}
  {\bibfnamefont{P.}~\bibnamefont{Senellart}}, \bibinfo {author}
  {\bibfnamefont{A.}~\bibnamefont{Lema{\^i}tre}}, \bibinfo {author}
  {\bibfnamefont{B.}~\bibnamefont{Sermage}},\ and\ \bibinfo {author}
  {\bibfnamefont{J.}~\bibnamefont{Bloch}},\ }%
  \bibfield{title}{%
  \enquote{\bibinfo {title} {Dynamics of microcavity polaritons in the presence
  of an electron gas},}\ }%
  \bibfield{journal}{%
  \Doi{10.1103/PhysRevB.73.205344}{\bibinfo {journal} {Phys. Rev. B}}\ }%
  \textbf{\bibinfo {volume} {73}},\ \bibinfo {pages} {205344} (\bibinfo {year}
  {2006})%
  \bibAnnoteFile{NoStop}{Bloch_DynamicsElectronGas}%
\bibitem{Bloch_PropagationAmplification}%
  \BibitemOpen
  \bibfield{author}{%
  \bibinfo {author} {\bibfnamefont{E.}~\bibnamefont{Wertz}}, \bibinfo {author}
  {\bibfnamefont{A.}~\bibnamefont{Amo}}, \bibinfo {author}
  {\bibfnamefont{D.~D.}\ \bibnamefont{Solnyshkov}}, \bibinfo {author}
  {\bibfnamefont{L.}~\bibnamefont{Ferrier}}, \bibinfo {author}
  {\bibfnamefont{T.~C.~H.}\ \bibnamefont{Liew}}, \bibinfo {author}
  {\bibfnamefont{D.}~\bibnamefont{Sanvitto}}, \bibinfo {author}
  {\bibfnamefont{P.}~\bibnamefont{Senellart}}, \bibinfo {author}
  {\bibfnamefont{I.}~\bibnamefont{Sagnes}}, \bibinfo {author}
  {\bibfnamefont{A.}~\bibnamefont{Lema\^{\i}tre}}, \bibinfo {author}
  {\bibfnamefont{A.~V.}\ \bibnamefont{Kavokin}}, \bibinfo {author}
  {\bibfnamefont{G.}~\bibnamefont{Malpuech}},\ and\ \bibinfo {author}
  {\bibfnamefont{J.}~\bibnamefont{Bloch}},\ }%
  \bibfield{title}{%
  \enquote{\bibinfo {title} {Propagation and amplification dynamics of 1d
  polariton condensates},}\ }%
  \bibfield{journal}{%
  \Doi{10.1103/PhysRevLett.109.216404}{\bibinfo {journal} {Phys. Rev. Lett.}}\
  }%
  \textbf{\bibinfo {volume} {109}},\ \bibinfo {pages} {216404} (\bibinfo {year}
  {2012})%
  \bibAnnoteFile{NoStop}{Bloch_PropagationAmplification}%
\bibitem{Gippius_Bistability}%
  \BibitemOpen
  \bibfield{author}{%
  \bibinfo {author} {\bibfnamefont{S.~S.}\ \bibnamefont{Gavrilov}}, \bibinfo
  {author} {\bibfnamefont{A.~S.}\ \bibnamefont{Brichkin}}, \bibinfo {author}
  {\bibfnamefont{A.~A.}\ \bibnamefont{Demenev}}, \bibinfo {author}
  {\bibfnamefont{A.~A.}\ \bibnamefont{Dorodnyy}}, \bibinfo {author}
  {\bibfnamefont{S.~I.}\ \bibnamefont{Novikov}}, \bibinfo {author}
  {\bibfnamefont{V.~D.}\ \bibnamefont{Kulakovskii}}, \bibinfo {author}
  {\bibfnamefont{S.~G.}\ \bibnamefont{Tikhodeev}},\ and\ \bibinfo {author}
  {\bibfnamefont{N.~A.}\ \bibnamefont{Gippius}},\ }%
  \bibfield{title}{%
  \enquote{\bibinfo {title} {Bistability and nonequilibrium transitions in the
  system of cavity polaritons under nanosecond-long resonant excitation},}\ }%
  \bibfield{journal}{%
  \Doi{10.1103/PhysRevB.85.075319}{\bibinfo {journal} {Phys. Rev. B}}\ }%
  \textbf{\bibinfo {volume} {85}},\ \bibinfo {pages} {075319} (\bibinfo {year}
  {2012})%
  \bibAnnoteFile{NoStop}{Gippius_Bistability}%
\bibitem{Deveaud_Josephson}%
  \BibitemOpen
  \bibfield{author}{%
  \bibinfo {author} {\bibfnamefont{K.~G.}\ \bibnamefont{Lagoudakis}}, \bibinfo
  {author} {\bibfnamefont{B.}~\bibnamefont{Pietka}}, \bibinfo {author}
  {\bibfnamefont{M.}~\bibnamefont{Wouters}}, \bibinfo {author}
  {\bibfnamefont{R.}~\bibnamefont{Andr\'e}},\ and\ \bibinfo {author}
  {\bibfnamefont{B.}~\bibnamefont{Deveaud-Pl\'edran}},\ }%
  \bibfield{title}{%
  \enquote{\bibinfo {title} {Coherent oscillations in an exciton-polariton
  josephson junction},}\ }%
  \bibfield{journal}{%
  \Doi{10.1103/PhysRevLett.105.120403}{\bibinfo {journal} {Phys. Rev. Lett.}}\
  }%
  \textbf{\bibinfo {volume} {105}},\ \bibinfo {pages} {120403} (\bibinfo {year}
  {2010})%
  \bibAnnoteFile{NoStop}{Deveaud_Josephson}%
\bibitem{Deveaud_VortexDynamics}%
  \BibitemOpen
  \bibfield{author}{%
  \bibinfo {author} {\bibfnamefont{K.~G.}\ \bibnamefont{Lagoudakis}}, \bibinfo
  {author} {\bibfnamefont{F.}~\bibnamefont{Manni}}, \bibinfo {author}
  {\bibfnamefont{B.}~\bibnamefont{Pietka}}, \bibinfo {author}
  {\bibfnamefont{M.}~\bibnamefont{Wouters}}, \bibinfo {author}
  {\bibfnamefont{T.~C.~H.}\ \bibnamefont{Liew}}, \bibinfo {author}
  {\bibfnamefont{V.}~\bibnamefont{Savona}}, \bibinfo {author}
  {\bibfnamefont{A.~V.}\ \bibnamefont{Kavokin}}, \bibinfo {author}
  {\bibfnamefont{R.}~\bibnamefont{Andr\'e}},\ and\ \bibinfo {author}
  {\bibfnamefont{B.}~\bibnamefont{Deveaud-Pl\'edran}},\ }%
  \bibfield{title}{%
  \enquote{\bibinfo {title} {Probing the dynamics of spontaneous quantum
  vortices in polariton superfluids},}\ }%
  \bibfield{journal}{%
  \Doi{10.1103/PhysRevLett.106.115301}{\bibinfo {journal} {Phys. Rev. Lett.}}\
  }%
  \textbf{\bibinfo {volume} {106}},\ \bibinfo {pages} {115301} (\bibinfo {year}
  {2011})%
  \bibAnnoteFile{NoStop}{Deveaud_VortexDynamics}%
\bibitem{Deveaud_Disorder}%
  \BibitemOpen
  \bibfield{author}{%
  \bibinfo {author} {\bibfnamefont{A.}~\bibnamefont{Baas}}, \bibinfo {author}
  {\bibfnamefont{K.~G.}\ \bibnamefont{Lagoudakis}}, \bibinfo {author}
  {\bibfnamefont{M.}~\bibnamefont{Richard}}, \bibinfo {author}
  {\bibfnamefont{R.}~\bibnamefont{Andr\'e}}, \bibinfo {author}
  {\bibfnamefont{Le~Si}\ \bibnamefont{Dang}},\ and\ \bibinfo {author}
  {\bibfnamefont{B.}~\bibnamefont{Deveaud-Pl\'edran}},\ }%
  \bibfield{title}{%
  \enquote{\bibinfo {title} {Synchronized and desynchronized phases of
  exciton-polariton condensates in the presence of disorder},}\ }%
  \bibfield{journal}{%
  \Doi{10.1103/PhysRevLett.100.170401}{\bibinfo {journal} {Phys. Rev. Lett.}}\
  }%
  \textbf{\bibinfo {volume} {100}},\ \bibinfo {pages} {170401} (\bibinfo {year}
  {2008})%
  \bibAnnoteFile{NoStop}{Deveaud_Disorder}%
\end{thebibliography}%

\end{document}